\documentclass[english,pra,preprint,showpacs]{revtex4}
\usepackage[T1]{fontenc}
\usepackage[latin1]{inputenc}
\usepackage{graphicx}

\makeatletter

\providecommand{\tabularnewline}{\\}

\usepackage{babel}
\makeatother
\begin{document}

\title{Kinetically-balanced Gaussian Basis Set Approach to Relativistic
Compton Profiles of Atoms}

\author{Prerit Jaiswal and Alok Shukla}

\email{shukla@phy.iitb.ac.in}

\affiliation{Physics Department, Indian Institute of Technology, Powai, Mumbai
400076, INDIA }

\begin{abstract}
Atomic Compton profiles (CPs) are a very important property which
provide us information about the momentum distribution of atomic electrons.
Therefore, for CPs of heavy atoms, relativistic effects are expected
to be important, warranting a relativistic treatment of the problem.
In this paper, we present an efficient approach aimed at \emph{ab
initio} calculations of atomic CPs within a Dirac-Hartree-Fock (DHF)
formalism, employing kinetically-balanced Gaussian basis functions.
The approach is used to compute the CPs of noble gases ranging from
He to Rn, and the results have been compared to the experimental and
other theoretical data, wherever possible. The influence of the quality
of the basis set on the calculated CPs has also been systematically
investigated.  
\end{abstract}


\maketitle

\section{Introduction}

\label{sec-intro}

Recent years have seen tremendous amount of progress in the field
of relativistic electronic structure calculations of atoms and molecules
using Dirac-equation-based approaches\cite{dirac-gen}. Particularly
noteworthy are the advances made in the field of basis-set-based relativistic
electronic structure theory pioneered by Kim\cite{kim}, and Kagawa\cite{kagawa}.
Although, initially, the basis-sets employed in the calculations were
of the ordinary Slater-type\cite{kim,kagawa}, however, now-a-days,
the preferred basis functions are those which incorporate the so-called
kinetic-balance condition between the large and the small component
basis functions\cite{ishikawa,dyall,kbgf}. The most commonly used
variety of such functions in relativistic electronic-structure calculations
are the kinetically-balanced Gaussian functions (KBGFs) which have
not only been instrumental in avoiding the problem of 'variational
collapse', but have also allowed the import of efficient algorithms
developed in basis-set-based nonrelativistic quantum chemistry. Using
such basis functions, calculations are now routinely performed both
at the mean-field Hartree-Fock (henceforth Dirac-Hartree-Fock (DHF))
level\cite{cal-at-dhf,cal-mol-dhf}, as well as at the correlated
level, employing methods such as the configuration-interaction (CI)
approach, both for atoms\cite{ci-shukla}, and molecules\cite{ci-relmol}. 

However, the progress in calculating wave functions and atomic energies
using KGBFs has not been matched by the progress in computing expectation
values corresponding to various physical quantities. For example,
atomic Compton profiles (CPs) are a very important property which
provide us information about the momentum distribution of atomic electrons,
and help us in interpreting the x-ray Compton scattering data from
atoms in the large momentum-transfer regime\cite{compton-review}.
Compton profiles are also very useful in understanding the bonding
properties, as one makes a transition from the atomic scale to the
scale of condensed matter\cite{compton-review}. Indeed, the nonrelativistic
Schr\"odinger equation based calculations of CPs of atomic and molecular
systems both within an \emph{ab initio}, as well as model-potential
based, formalisms are quite well developed\cite{compton-review}.
As recently demonstrated by us, and several other authors earlier
on, that such nonrelativistic \emph{ab initio} calculations of CPs
can also be performed on crystalline systems\cite{wannier}. However,
for systems involving heavy atoms, on intuitive grounds one expects
that the relativistic effects will become quite important, thereby
requiring a relativistic treatment of the problem\cite{kane-review}.
Long time back Mendelsohn \emph{et al.}\cite{rel-com-prof}, and Bigss
\emph{et al.}\cite{rel-comp-prof-2} presented the first fully-relativistic
calculations of atomic CPs which were performed at the DHF level,
employing a finite-difference based numerical approach. Yet, since
that time, there has been hardly any activity in the field, which
is surprising given the fact that now relativistic electronic structure
calculations are routinely performed employing KBGF basis functions.
Therefore, in this work, our aim is to report the first calculations
of atomic CPs at the DHF level, employing a basis set composed of
KBGFs. Our approach is based upon analytic formulas for the CP matrix
elements with respect to a KBGF basis set, whose derivation is presented
in the Appendix. The DHF calculations of atomic CPs are presented
for the entire rare gas series (He to Rn), and our results are compared
to experimental data, wherever available. Additionally, our results
for Ar, Kr, Xe, and Rn are also compared to the DHF results of Mendelsohn
\emph{et al.}\cite{rel-com-prof}, and Bigss \emph{et al.}\cite{rel-comp-prof-2},
and excellent agreement is obtained between the two sets of calculations. 

At this point we would like to clarify one important aspect related
to the relativistic effects which our calculations are computing,
in light of the fact that there have been several papers in the literature
dealing with a relativistic treatment of Compton scattering of bound
electrons\cite{ribber,holm,berg}. Several authors have pointed out
that for very large photon energies, a fully relativistic treatment,
within the framework of quantum-electrodynamics, of the Compton scattering
from bound electrons is essential\cite{kane-review}. When such a
treatment of the problem is performed, it is not clear whether the
Compton scattering cross-sections can at all be written in terms of
Compton profiles\cite{ribber,holm,kane-review,berg}. Our work presented
here, however, does not correspond to that regime of photon energies.
What we mean by the relativistic effects here are the changes in the
computed CPs because of a relativistic treatment of the bound electrons
within a Dirac Hamiltonian based formalism. Thus, our calculations
assume that the Compton scattering from atomic electrons can be described
in terms of the CPs under the impulse approximation\cite{impuse}.
The electron momentum densities needed to calculate the CPs, however,
are computed from the Dirac orbitals of the atomic electrons. This
approach is identical to the one adopted in the earlier DHF calculations\cite{rel-com-prof,rel-comp-prof-2}. 

The remainder of this paper is organized as follows. In section \ref{sec-theory}
we present the basic theoretical formalism behind the present set
of calculations. Next in section \ref{sec-results} we present and
discuss the results of our calculations. Finally, in section \ref{sec-conclusions}
our conclusions, as well as possible future directions for further
work are discussed. Additionally, in the Appendix we present the derivation
of the closed-form formulas for CPs over KBGFs, used in our calculations.

\section{Theory}

\label{sec-theory}

Our theory is based upon the Dirac-Coulomb Hamiltonian\begin{equation}
H=\sum_{i}(c{\bf \alpha}_{i}\cdot{\bf p}_{i}+c^{2}(\beta_{i}-1)+V_{\mbox{nuc}}(r_{i}))+\sum_{i<j}\frac{1}{r_{ij}},\label{eq-dirac}\end{equation}
where $c$ is the speed of light, ${\bf p}$ is the momentum operator,
$V_{\mbox{nuc}}(r)$ is the electron-nucleus interaction potential,
indices $i$ and $j$ label the electrons of the atom, and $r_{ij}$
is the distance between the $i$th and $j$th electrons. For $V_{\mbox{nuc}}(r)$
a spherical finite-nucleus approximation is employed, with the radius
estimated as $2.2677\times10^{-5}A^{1/3}$, where $A$ is the atomic
mass number\cite{cal-at-dhf}. The Dirac matrices are chosen to be
${\bf \alpha}=\left(\begin{array}{cc}
{\bf 0} & {\bf \sigma}\\
{\bf \sigma} & {\bf {\bf 0}}\end{array}\right)$ and $\beta=\left(\begin{array}{ll}
{\bf I} & {\bf 0}\\
{\bf 0} & {\bf -I}\end{array}\right)$, where ${\bf 0}$, ${\bf I}$, and ${\bf \sigma}$, represent the
2$\times$2 null, identity and Pauli matrices, respectively. Eq. (\ref{eq-dirac})
is solved under the DHF approximation utilizing spherical symmetry
with the orbitals of the form\begin{equation}
\psi_{n\kappa m}=r^{-1}\left(\begin{array}{l}
P_{n\kappa}(r)\chi_{\kappa m}(\theta,\phi)\\
iQ_{n\kappa}(r)\chi_{-\kappa m}(\theta,\phi)\end{array}\right),\label{eq-spinor}\end{equation}
where $P_{n\kappa}(r)$ and $Q_{n\kappa}(r)$ are the radial large
and small components, and $\chi_{\kappa m}(\theta,\phi)$ is the two-component
angular part composed of Clebsch-Gordon coefficients and spherical
harmonics. In the basis-set approach adopted here, the radial parts
of the wave function are expressed as linear combination of radial
Gaussian type of functions\[
P_{n\kappa}(r)=\sum_{i}C_{\kappa i}^{L}g_{\kappa i}^{L}(r),\]
and \[
Q_{n\kappa}(r)=\sum_{i}C_{\kappa i}^{S}g_{\kappa i}^{S}(r),\]
where $C_{\kappa i}^{L}$ and $C_{\kappa i}^{S}$, are the expansion
coefficients of the large and small component basis functions, respectively.
The large-component basis function is given by\begin{equation}
g_{\kappa i}^{L}(r)=N_{\kappa i}^{L}r^{n_{\kappa}}e^{-\alpha_{i}r^{2}},\label{eq-glarge}\end{equation}
while the small-component basis function is obtained by the kinetic-balancing
condition\cite{kbgf}\begin{equation}
g_{\kappa i}^{S}=N_{\kappa i}^{S}(\frac{d}{dr}+\frac{\kappa}{r})g_{\kappa i}^{L}(r).\label{eq-gsmall}\end{equation}
 Above $n_{\kappa}$ is the principal quantum number associated with
a symmetry species ($n_{\kappa}=1,2,2,3,3,\ldots$, for symmetry species
$s$, $p_{1/2}$, $p_{3/2}$, $d_{3/2}$, $d_{5/2}$, $\ldots$),
$\alpha_{i}$ is the Gaussian exponent of the $i$th basis function,
and $N_{\kappa i}^{L}$, $N_{\kappa i}^{S}$ are the normalization
coefficients associated with the large and the small component basis
functions, respectively.

Under the impulse approximation\cite{impuse}, the differential cross-section
of Compton scattering of x-rays from many-electron systems is proportional
to the Compton profile\begin{equation}
J(q)=\int\int dp_{x}dp_{y}\rho({\bf p}),\label{eq-cp}\end{equation}
where $\rho({\bf p})$ is the momentum distribution of the electrons
before scattering and $q$ is the component of the momentum of the
electron along the scattering vector, assumed to be along the $z$
direction. Under the mean-field DHF approximation, for a closed-shell
atom, the expression for the CP reduces to\begin{equation}
J(q)=\sum(2j_{i}+1)J_{n_{i}\kappa_{i}}(q),\label{eq-jq}\end{equation}
where $j_{i}$ is the total angular momentum of the $i$th orbital
while, $J_{n_{i}\kappa_{i}}$ is the CP associated with it \begin{equation}
J_{n_{i}\kappa_{i}}(q)=\frac{1}{2}\int_{q}^{\infty}\{|P_{n_{i}\kappa_{i}}(p)|^{2}+|Q_{n_{i}\kappa_{i}}(p)|^{2}\} pdp,\label{eq-jorb}\end{equation}
where $P_{n_{i}\kappa_{i}}(p)$ and $Q_{n_{i}\kappa_{i}}(p)$ are
the Fourier transforms of the radial parts of the large and small
components, respectively, of the $i$th occupied orbital (cf. Eq.
(\ref{eq-spinor})) and are defined as\begin{equation}
P_{n_{i}\kappa_{i}}(p)=\frac{4\pi}{(2\pi)^{3/2}}\int_{0}^{\infty}rP_{n_{i}\kappa_{i}}(r)j_{l_{A}}(pr)dr,\label{eq-pnk}\end{equation}
and\begin{equation}
Q_{n_{i}\kappa_{i}}(p)=\frac{4\pi}{(2\pi)^{3/2}}\int_{0}^{\infty}rQ_{n_{i}\kappa_{i}}(r)j_{l_{B}}(pr)dr,\label{eq-qnk}\end{equation}
where $j_{l_{A}}(pr)$($j_{l_{B}}(pr)$) is the spherical Bessel function
corresponding to the orbital angular momentum $l_{A}$($l_{B}$) of
the large (small) component. Therefore, calculation of atomic CPs
involves computation of two types of integrals: (i) radial Fourier
Transforms of Eqs. (\ref{eq-pnk}) and (\ref{eq-qnk}), and (ii) momentum
integrals of the Fourier transformed orbitals in Eq. (\ref{eq-jorb}).
When one solves the DHF equation for atoms using the finite-difference
techniques, then, obviously the calculation of atomic CPs mandates
that both these types integrals be computed by means of numerical
quadrature. However, for the basis-set-based approach adopted here,
in order to facilitate rapid computation of atomic CPs, it is desirable
to obtain closed-form expressions for both types of integrals with
respect to the chosen basis functions. Indeed, we have managed to
derive closed-form expressions for the atomic CPs with respect to
the KBGFs, which can be easily computer implemented. It is easy to
see that within a KBGF based approach, the integral of Eq. (\ref{eq-jorb}),
can be computed in terms of the following two types of integrals\[
J_{ij}^{L;\kappa}(q)=\frac{1}{2}\int_{q}^{\infty}pg_{\kappa i}^{L}(p)g_{\kappa j}^{L}(p)dp,\]
and \[
J_{ij}^{S;\kappa}(q)=\frac{1}{2}\int_{q}^{\infty}pg_{\kappa i}^{S}(p)g_{\kappa j}^{S}(p)dp,\]
 where $g_{\kappa i}^{L}(p)$ and $g_{\kappa i}^{S}(p)$ are the radial
Fourier Transforms (cf. Eqs. (\ref{eq-pnk}) and (\ref{eq-qnk}))
of the large and small component basis functions $g_{\kappa i}^{L}(r)$,
and $g_{\kappa i}^{S}(r)$, respectively. Obtaining closed-form expressions
for $J_{ij}^{L;\kappa}(q)$ and $J_{ij}^{S;\kappa}(q)$ expressions
was not an easy task, and those formulas, along with their derivation,
are presented in the Appendix. Additionally, elsewhere we have described
a Fortran 90 computer program developed by us, which uses these expressions
to compute the atomic CPs from a set of given Dirac orbitals expressed
as a linear combination of KBGFs\cite{cp-program}.

Here we would like to comment on possible quantitative manifestations
of relativistic effects in Compton profiles. One obvious way to quantify
the relativistic effects on the CPs is by comparing the values obtained
from the DHF calculations with those obtained from nonrelativistic
HF calculations. There is another way by which one can judge the influence
of relativistic effects on Compton profiles, that is by comparing
the orbital CPs of different fine structure components. For example,
in nonrelativistic calculations, $np$, $nd$,$\dots$ orbitals have
only one set of values each for the orbital CPs. However, in relativistic
calculations, each such orbital splits into two fine-structure components,
\emph{i.e.}, $np_{1/2}/np_{3/2}$, $nd_{3/2}/nd_{5/2}$, which, if
the relativistic effects are strong, can differ from each other in
a significant manner. Thus, one expects, that under such situations,
the orbital profiles of the two fine-structure components will also
be significantly different. Therefore, we will also examine this ``fine-structure
splitting'' of the orbital CPs of various atoms to quantify the relativistic
effects.

\section{Calculations and Results}

\label{sec-results}

In this section we present our DHF results on the atomic profiles
of the rare gases. The DHF orbitals of various atoms were computed
using the KBGF based REATOM code of Mohanty and Clementi\cite{reatom}.
During the DHF calculations the value of the speed of light used was
$c=137.037$ a.u. Additionally, for obtaining the radius of the nucleus
for the finite-nucleus approximation description of $V_{\mbox{nuc}}(r)$,
values of atomic mass $A$ were taken to be $4.026$, $20.18$, $39.948$,
$83.80$, $131.3$, and $222.0$ for He, Ne, Ar, Kr, Xe, and Rn, respectively.
Using the orbitals obtained from the DHF calculations, the atomic
CPs were computed using our computer program COMPTON\cite{cp-program}.
Next we present our results for the rare gas atoms, one-by-one. In
order to investigate the basis-set dependence of the CPs, for each
atom, two types of basis sets were used: (i) a large universal basis
set proposed by Malli \emph{et al.} \cite{basis-malli}\emph{,} and
(ii) a smaller basis set tailor-made for the individual atom.

\subsection{He }

\label{results-he}

\begin{figure}
\includegraphics[width=10cm,keepaspectratio]{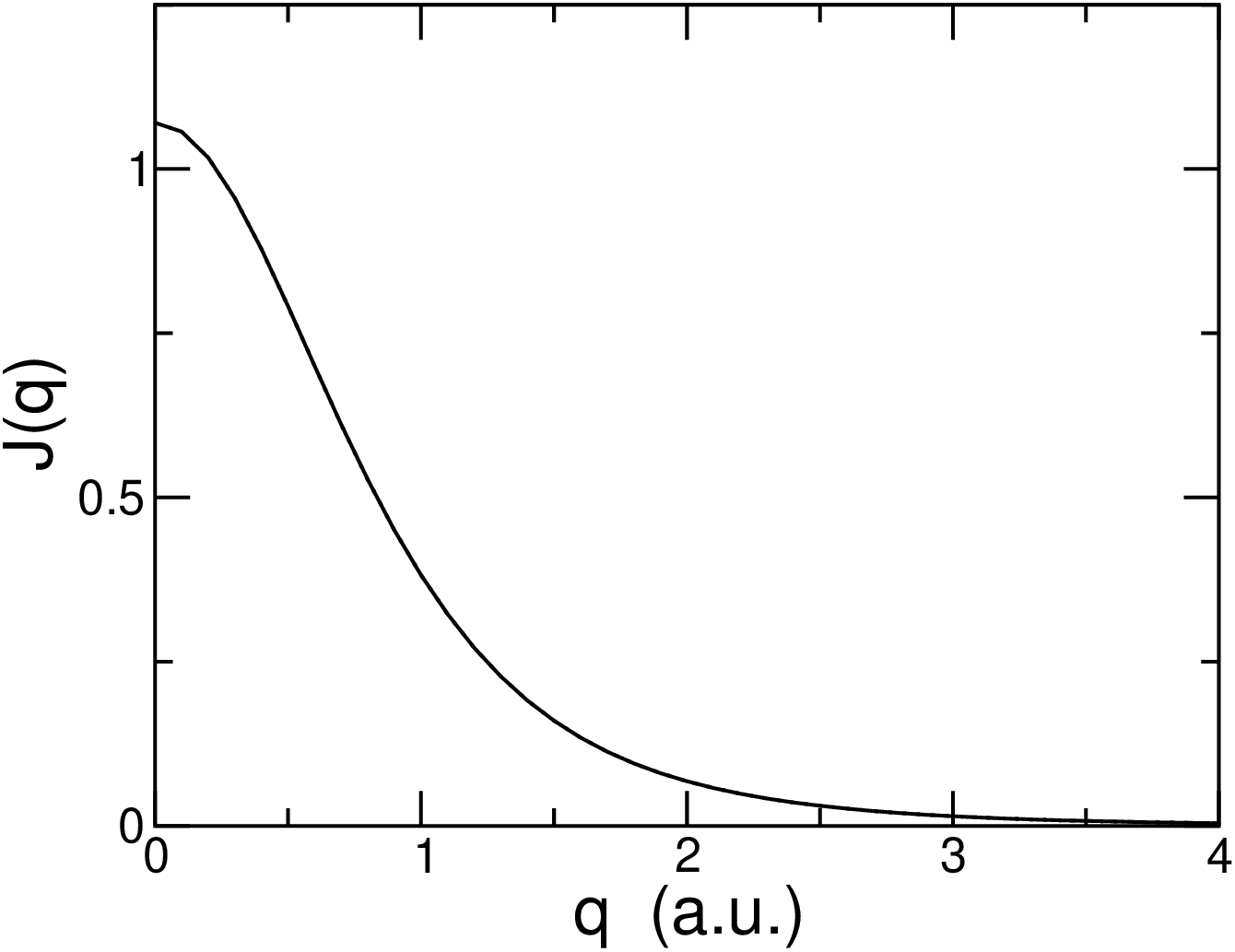}\label{fig-hecp}

\caption{DHF Compton profiles of He, $J(q)$, computed using the well-tempered
basis set\cite{basis-mat}, and the universal basis set\cite{basis-malli},
as a function of the momentum transfer $q$. Profiles obtained using
the two basis functions are virtually indistinguishable. }
\end{figure}

For He, DHF calculations were performed with: (i) well-tempered basis
set of Matsuoka and Huzinaga\cite{basis-mat} employing $12s$ functions\cite{basis-mat},
and (ii) the universal basis set using $22s$ functions\cite{basis-malli}.
The computed CPs are plotted in Fig. \ref{fig-hecp} as a function
of the momentum transfer $q$. The results of our calculations for
some selected values of $q$ are presented in table \ref{tab-he}.
For the sake of comparison, the same table also contains the nonrelativistic
HF results of Clementi and Roetti\cite{he-hf}, as well as the experimental
results of Eisenberger and Reed\cite{cp-exp-1}. Upon inspection of
the table, following trends emerge: (i) Our relativistic CPs computed
with the well-tempered and the universal basis sets are in excellent
agreement with each other. This implies that the smaller well-tempered
basis set is virtually complete, as far as the CPs are concerned.
(ii) Our DHF CPs are in excellent agreement with the nonrelativistic
HF CPs of Clementi and Roetti\cite{he-hf}. This, obviously, is a
consequence of the fact that the relativistic effects are negligible
for a light atom such as He. (iii) Generally, the agreement between
the theoretical and the experimental CPs is excellent, implying that
the electron-correlation effects do not make a significant contribution
in this case.

\begin{table}

\caption{Relativistic (DHF) Compton profiles of He atom computed using various
basis functions, compared to the nonrelativistic HF results\cite{he-hf},
and the experiments\cite{cp-exp-1}.   }

\begin{tabular}{|r|c|c|c|c|}
\hline 
$q$(a.u.)&
$J(q)$(WT)$^{a}$&
$J(q)$(Uni)$^{b}$&
$J(q)$(HF)$^{c}$&
$J(q)$(Exp.)$^{d}$\tabularnewline
\hline
\hline 
$0.0$&
$1.0704$&
$1.0704$&
$1.0705$&
$1.071\pm1.5\%$\tabularnewline
\hline 
$0.1$&
$1.0567$&
$1.0567$&
$1.0568$&
$1.058$\tabularnewline
\hline 
$0.2$&
$1.0171$&
$1.0171$&
$1.017$&
$1.019$\tabularnewline
\hline 
$0.3$&
$0.9557$&
$0.9557$&
$0.955$&
$0.958$\tabularnewline
\hline 
$0.4$&
$0.8782$&
$0.8782$&
$0.878$&
$0.881$\tabularnewline
\hline 
$0.5$&
$0.7910$&
$0.7910$&
$0.791$&
$0.795$\tabularnewline
\hline 
$0.6$&
$0.7003$&
$0.7004$&
$0.700$&
$0.705$\tabularnewline
\hline 
$0.7$&
$0.6111$&
$0.6112$&
$0.611$&
$0.616$\tabularnewline
\hline 
$0.8$&
$0.5270$&
$0.5270$&
$0.527$&
$0.533\pm2.3\%$\tabularnewline
\hline 
$0.9$&
$0.4503$&
$0.4503$&
$0.450$&
$0.456$\tabularnewline
\hline 
$1.0$&
$0.3820$&
$0.3820$&
$0.382$&
$0.388$\tabularnewline
\hline 
$1.2$&
$0.2712$&
$0.2712$&
$0.271$&
$0.274$\tabularnewline
\hline 
$1.4$&
$0.1910$&
$0.1910$&
$0.190$&
$0.188$\tabularnewline
\hline 
$1.6$&
$0.1344$&
$0.1345$&
$0.134$&
$0.129$\tabularnewline
\hline 
$1.8$&
$0.0952$&
$0.0952$&
$0.095$&
$0.092$\tabularnewline
\hline 
$2.0$&
$0.0678$&
$0.0678$&
$0.068$&
$0.069$\tabularnewline
\hline 
$2.5$&
$0.0307$&
$0.0307$&
$0.031$&
$0.030\pm15\%$\tabularnewline
\hline 
$3.0$&
$0.0148$&
$0.0148$&
$0.015$&
$0.013$\tabularnewline
\hline 
$5.0$&
$0.0014$&
$0.0014$&
---&
---\tabularnewline
\hline 
$8.0$&
$0.0001$&
$0.0001$&
---&
---\tabularnewline
\hline 
$10.0$&
$0.00003$&
$0.00003$&
---&
---\tabularnewline
\hline
\end{tabular}\label{tab-he}

\begin{raggedright}
$^{a}$Our DHF results computed using the well-tempered basis set\cite{basis-mat}
\par\end{raggedright}

\begin{raggedright}
$^{b}$Our DHF results computed using the universal basis set\cite{basis-malli}
\par\end{raggedright}

\begin{raggedright}
$^{c}$Nonrelativistic HF results from Ref.\cite{he-hf}
\par\end{raggedright}

\begin{raggedright}
$^{d}$Experimental results from Ref.\cite{cp-exp-1}
\par\end{raggedright}
\end{table}

\subsection{Ne}

\label{results-ne}

\begin{figure}
\includegraphics[width=10cm,keepaspectratio]{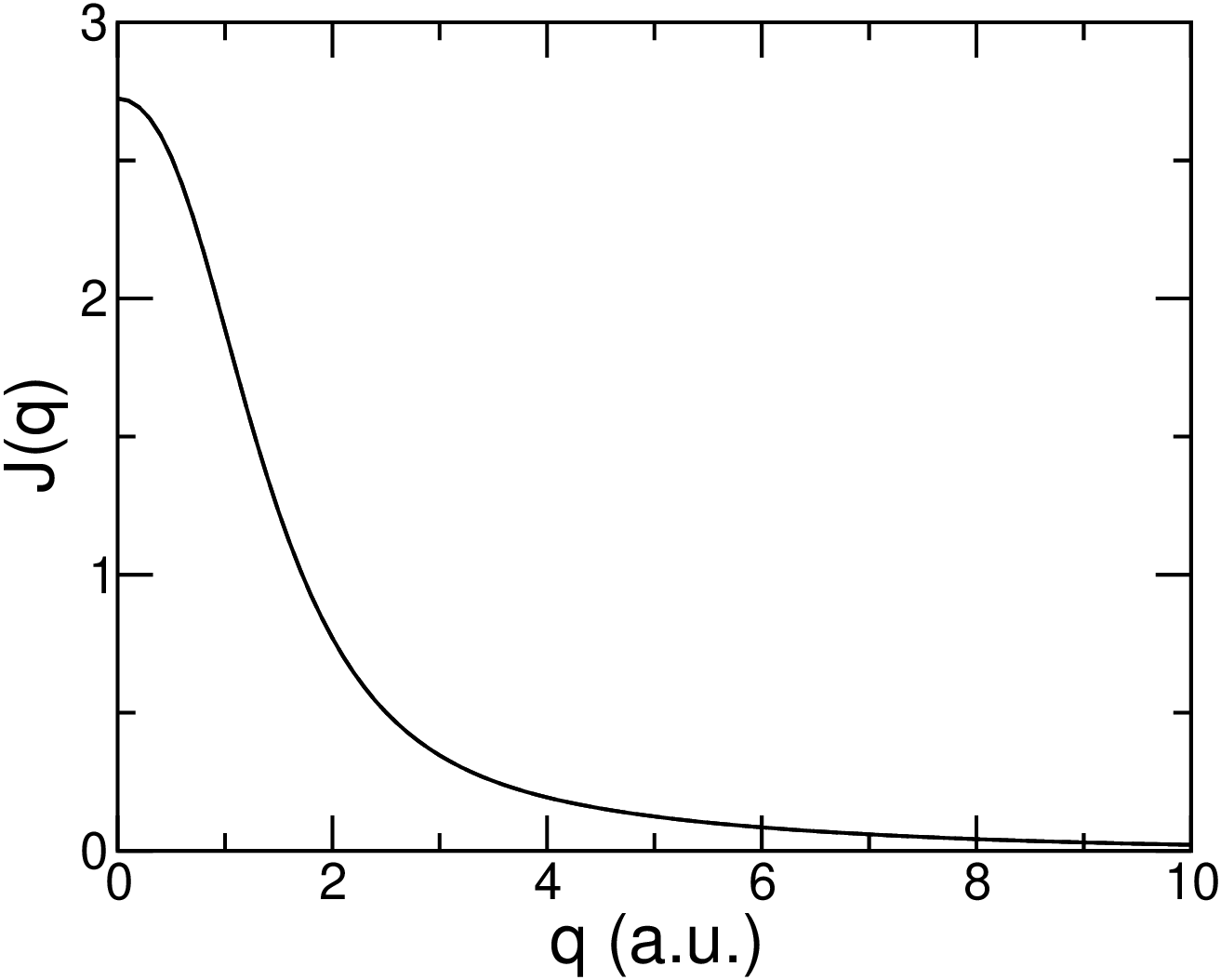}

\caption{DHF Compton profiles of Ne, $J(q)$, computed using the well-tempered
basis set\cite{basis-mat}, and the universal basis set\cite{basis-malli},
as a function of the momentum transfer $q$. Profiles obtained using
the two basis sets are virtually indistinguishable. All numbers are
in atomic units.}

\label{fig-ne-cp}
\end{figure}

DHF calculations were performed for Ne using: (i) \textbf{\large }$(14s$,
$14p$) well-tempered basis set of Matsuoka and Huzinaga\cite{basis-mat},
and the (ii) large ($32s$, $29p$) universal basis set of Malli \emph{et
al.}\cite{basis-malli}\emph{.} In order to facilitate direct comparison
with the experiments, the valence CPs (excluding the contribution
from the $1s$ core orbital) obtained from our calculations are presented
in table \ref{tab-ne}. They are also compared to the nonrelativistic
HF results of Clementi and Roetti\cite{he-hf}, classic experiment
of Eisenberger\cite{exp-ne-eis}, and more recent experiment of Lahmam-Bennani
\emph{et al.}\cite{exp-ne-jcp}. Additionally, the total Compton profiles
of Ne (including the contribution of the $1s$ orbital), computed
using both the aforesaid basis sets, are plotted in Fig. \ref{fig-ne-cp}.

\begin{table}

\caption{Relativistic (DHF) valence Compton profiles of Ne atom computed using
various basis functions, compared to the nonrelativistic HF results\cite{he-hf},
and the experiments\cite{cp-exp-1}.  }

\begin{tabular}{|c|c|c|c|c|c|}
\hline 
$q$(a.u.)&
$J(q)$(WT)$^{a}$&
$J(q)$(Uni)$^{b}$&
$J(q)$ (HF)$^{c}$&
$J(q)$(Exp.)$^{d}$&
$J(q)$(Exp.)$^{e}$\tabularnewline
\hline
\hline 
$0.0$&
$2.5439$&
$2.5452$&
$2.548$&
$2.582$&
$2.602$\tabularnewline
\hline 
$0.1$&
$2.5363$&
$2.5375$&
$2.540$&
$2.574$&
$2.593$\tabularnewline
\hline 
$0.2$&
$2.5128$&
$2.5140$&
$2.515$&
$2.558$&
$2.560$\tabularnewline
\hline 
$0.3$&
$2.4722$&
$2.4731$&
$2.475$&
$2.519$&
$2.506$\tabularnewline
\hline
$0.4$&
$2.4129$&
$2.4133$&
$2.418$&
$2.451$&
$2.435$\tabularnewline
\hline
$0.5$&
$2.3342$&
$2.3339$&
$2.335$&
$2.359$&
$2.340$\tabularnewline
\hline
$0.6$&
$2.2367$&
$2.2357$&
$2.236$&
$2.249$&
$2.235$\tabularnewline
\hline
$0.7$&
$2.1224$&
$2.1210$&
$2.120$&
$2.124$&
$2.099$\tabularnewline
\hline
$0.8$&
$1.9947$&
$1.9933$&
$1.990$&
$1.986$&
$1.966$\tabularnewline
\hline
$0.9$&
$1.8579$&
$1.8568$&
$1.855$&
$1.839$&
$1.826$\tabularnewline
\hline
$1.0$&
$1.7166$&
$1.7159$&
$1.715$&
$1.685$&
$1.690$\tabularnewline
\hline
$1.2$&
$1.4360$&
$1.4361$&
$1.435$&
$1.394$&
$1.417$\tabularnewline
\hline
$1.4$&
$1.1776$&
$1.1780$&
$1.171$&
$1.140$&
$1.171$\tabularnewline
\hline
$1.6$&
$0.9533$&
$0.9537$&
$0.951$&
$0.921$&
$0.975$\tabularnewline
\hline
$1.8$&
$0.7663$&
$0.7665$&
$0.766$&
$0.749$&
---\tabularnewline
\hline
$2.0$&
$0.6142$&
$0.6144$&
$0.619$&
$0.608$&
---\tabularnewline
\hline
$2.5$&
$0.3559$&
$0.3558$&
$0.355$&
$0.355$&
---\tabularnewline
\hline
$3.0$&
$0.2125$&
$0.2123$&
$0.212$&
$0.225$&
---\tabularnewline
\hline
$3.5$&
$0.1318$&
$0.1319$&
$0.132$&
$0.156$&
---\tabularnewline
\hline
$4.0$&
$0.0852$&
$0.0853$&
$0.085$&
$0.102$&
---\tabularnewline
\hline
$5.0$&
$0.0397$&
$0.0397$&
$0.040$&
$0.041$&
---\tabularnewline
\hline
\end{tabular}

\begin{raggedright}
$^{a}$Our DHF results computed using the well-tempered basis set\cite{basis-mat}
\par\end{raggedright}

\begin{raggedright}
$^{b}$Our DHF results computed using the universal basis set\cite{basis-malli}
\par\end{raggedright}

\begin{raggedright}
$^{c}$Nonrelativistic HF results from Ref.\cite{he-hf}
\par\end{raggedright}

\begin{raggedright}
$^{d}$Experimental results from Ref.\cite{exp-ne-eis}\\
$^{e}$Experimental results from Ref.\cite{exp-ne-jcp}
\par\end{raggedright}

\label{tab-ne}
\end{table}

Upon inspecting table \ref{tab-ne} we notice the following trends:
(i) profiles computed using two different sets are again in very good
agreement with each other, implying that both the basis sets are essentially
complete, (ii) our relativistic profiles are in quite good agreement
with the nonrelativistic HF profiles\cite{he-hf} essentially implying
that even in Ne, the relativistic effects are quite negligible. As
far as comparison with the experiments is concerned, for smaller values
of $q$ there is slight disagreement with the theory which progressively
disappears as one approaches the large momentum-transfer regime. This
suggests that electron-correlation effects possibly play an important
role in the small momentum transfer regime.

\begin{figure}
\includegraphics[width=10cm,keepaspectratio]{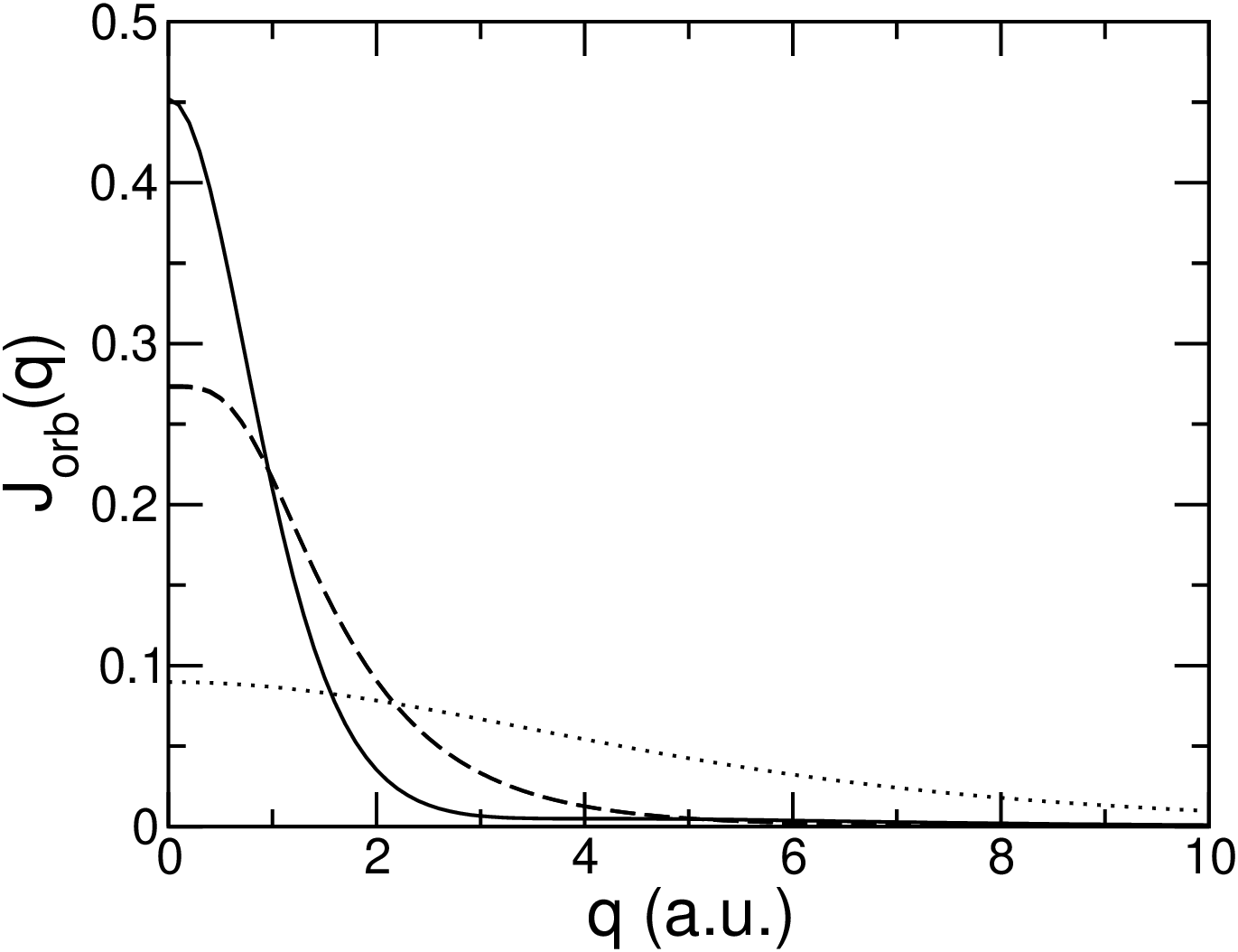}

\caption{Orbital Compton profiles of Ne for $2s$ (solid line), $2p_{3/2}$/$2p_{1/2}$
(dashed line), and $1s$ (dotted line), plotted with respect to $q.$
Compton profiles of $2p_{1/2}$ and $2p_{3/2}$ orbitals are virtually
indistinguishable. These profiles were computed using the universal
basis set\cite{basis-malli}.}

\begin{raggedright}
\label{fig-ne-orb-cp}
\par\end{raggedright}
\end{figure}

Finally we examine the individual orbital CPs of the Ne atom in Fig.
\ref{fig-ne-orb-cp}. The maximum contribution to the total CP for
small values of momentum transfer comes from the $2s$ orbital, while
in the same region, the smallest contribution comes from the $1s$
core orbital. The orbital CP of the $2s$ orbital varies rapidly with
respect to $q$ and becomes quite small for $q\geq2$ a.u. On the
other hand the orbital profile of the $1s$ orbital shows the least
dispersion with respect to $q$, and has the largest magnitude in
the large $q$ region, as compared to other orbital profiles. The
behavior of the $2p_{1/2}/2p_{3/2}$ orbital profiles is intermediate
as compared to the two extremes of $1s$ and $2s$ profiles. These
profiles have lesser magnitude compared to the $2s$ profile for $q\approx0$,
while they vary more rapidly with respect to $q$, when compared to
the $1s$ profile. Another pointer to the insignificance of the relativistic
effects for Ne is the fact that the difference in the values of the
$2p_{1/2}$ and $2p_{3/2}$ is quite small for all values of $q$. 

\subsection{Ar}

\label{sec-Ar}

Next, we discuss our calculated Compton profiles of Ar. The DHF calculations
on Ar atom were performed using the following two basis sets: (i)
smaller $(16s$,$16p$) well-tempered basis set of  Matsuoka and Huzinaga\cite{basis-mat},
and the (ii) large ($32s$,$29p)$ universal basis set of Malli \emph{et
al.}\cite{basis-malli}. Calculated total CPs of Ar, for a selected
number of $q$ values in the range $0$ a.u.$\leq q\leq15$ a.u.,
are presented in table \ref{tab-ar}. The same table also contains
the nonrelativistic HF results of Clementi and Roetti\cite{he-hf},
numerical-orbital-based DHF results of Mendelsohn \emph{et al.}\cite{rel-com-prof},
and the experimental results of Eisenberger and Reed\cite{cp-exp-1}.

\begin{table}

\caption{Our relativistic (DHF) total Compton profiles of Ar atom computed
using various basis sets, compared to the relativistic results of
other authors\cite{rel-com-prof}, the nonrelativistic HF results\cite{he-hf},
and the experiments\cite{cp-exp-1}.  }

\begin{tabular}{|c|c|c|c|c|c|}
\hline 
$q$(a.u.)&
$J(q)$(WT)$^{a}$&
$J(q)$(Uni)$^{b}$&
$J(q)$(DHF)$^{c}$&
$J(q)$(HF)$^{d}$&
$J(q)$(Exp.)$^{e}$\tabularnewline
\hline
\hline 
$0.0$&
$5.0471$&
$5.0543$&
$5.05$&
$5.052$&
$5.058\pm0.7\%$\tabularnewline
\hline 
$0.1$&
$5.0229$&
$5.0302$&
$5.03$&
$5.028$&
$5.022$\tabularnewline
\hline 
$0.2$&
$4.9473$&
$4.9539$&
$4.95$&
$4.950$&
$4.917$\tabularnewline
\hline 
$0.3$&
$4.8130$&
$4.8171$&
---&
$4.812$&
$4.749$\tabularnewline
\hline 
$0.4$&
$4.6143$&
$4.6144$&
$4.61$&
$4.608$&
$4.526$\tabularnewline
\hline 
$0.5$&
$4.3528$&
$4.3487$&
---&
$4.369$&
$4.259$\tabularnewline
\hline 
$0.6$&
$4.0395$&
$4.0324$&
$4.03$&
$4.028$&
$3.960$\tabularnewline
\hline 
$0.7$&
$3.6928$&
$3.6854$&
---&
$3.690$&
$3.643$\tabularnewline
\hline 
$0.8$&
$3.3343$&
$3.3288$&
---&
$3.328$&
$3.319$\tabularnewline
\hline 
$0.9$&
$2.9842$&
$2.9814$&
---&
$2.982$&
$3.000$\tabularnewline
\hline 
$1.0$&
$2.6576$&
$2.6573$&
$2.66$&
$2.658$&
$2.697\pm1\%$\tabularnewline
\hline 
$1.2$&
$2.1071$&
$2.1088$&
---&
$2.108$&
$2.164$\tabularnewline
\hline 
$1.4$&
$1.7011$&
$1.7022$&
---&
$1.701$&
$1.753$\tabularnewline
\hline 
$1.6$&
$1.4163$&
$1.4166$&
---&
$1.417$&
$1.461$\tabularnewline
\hline 
$1.8$&
$1.2198$&
$1.2197$&
---&
$1.221$&
$1.264$\tabularnewline
\hline 
$2.0$&
$1.0825$&
$1.0824$&
$1.08$&
$1.084$&
$1.129$\tabularnewline
\hline
$2.5$&
$0.8728$&
$0.8727$&
---&
$0.873$&
$0.904$\tabularnewline
\hline
$3.0$&
$0.7360$&
$0.7360$&
---&
$0.736$&
$0.744$\tabularnewline
\hline
$3.5$&
$0.6216$&
$0.6217$&
---&
$0.621$&
$0.634$\tabularnewline
\hline
$4.0$&
$0.5207$&
$0.5208$&
$0.521$&
$0.520$&
$0.534\pm2.5\%$\tabularnewline
\hline
$7.0$&
$0.1773$&
$0.1774$&
---&
$0.177$&
$0.181$\tabularnewline
\hline
$8.0$&
$0.1300$&
$0.1300$&
---&
$0.130$&
$0.137$\tabularnewline
\hline
$9.0$&
$0.0981$&
$0.0981$&
---&
$0.098$&
$0.104$\tabularnewline
\hline
$10.0$&
$0.0758$&
$0.0757$&
$0.076$&
$0.075$&
$0.078\pm10\%$\tabularnewline
\hline
$15.0$&
$0.0254$&
$0.0254$&
---&
$0.025$&
$0.025$\tabularnewline
\hline
\end{tabular}

\begin{raggedright}
$^{a}$our DHF results computed using the well-tempered basis set\cite{basis-mat}
\par\end{raggedright}

\begin{raggedright}
$^{b}$our DHF results computed using the universal basis set\cite{basis-malli}
\par\end{raggedright}

\begin{raggedright}
$^{c}$DHF results of Mendelsohn \emph{et al.}\cite{rel-com-prof}
based upon finite-difference calculations
\par\end{raggedright}

\begin{raggedright}
$^{d}$Nonrelativistic HF results from Ref.\cite{he-hf}
\par\end{raggedright}

\begin{raggedright}
$^{e}$Experimental results from Ref.\cite{cp-exp-1}
\par\end{raggedright}

\label{tab-ar}
\end{table}

\begin{figure}
\includegraphics[width=10cm,keepaspectratio]{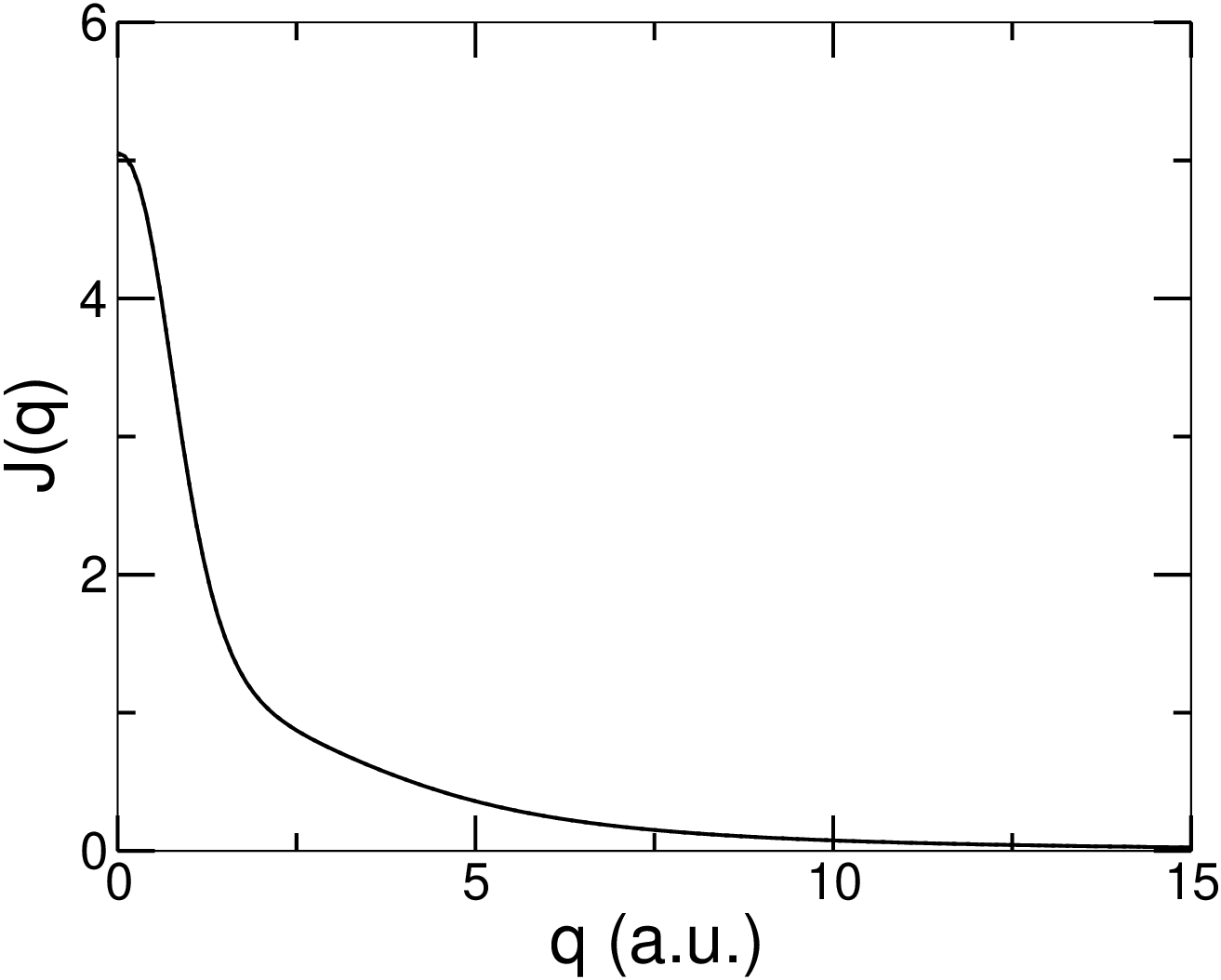}

\caption{DHF Compton profiles of Ar, $J(q)$, computed using the well-tempered
basis set\cite{basis-mat}, and the universal basis set\cite{basis-malli},
as a function of the momentum transfer $q$. Profiles obtained using
the two basis sets can be seen to differ slightly for $q\approx0$. }

\label{fig-ar}
\end{figure}

\begin{figure}
\includegraphics[width=10cm,keepaspectratio]{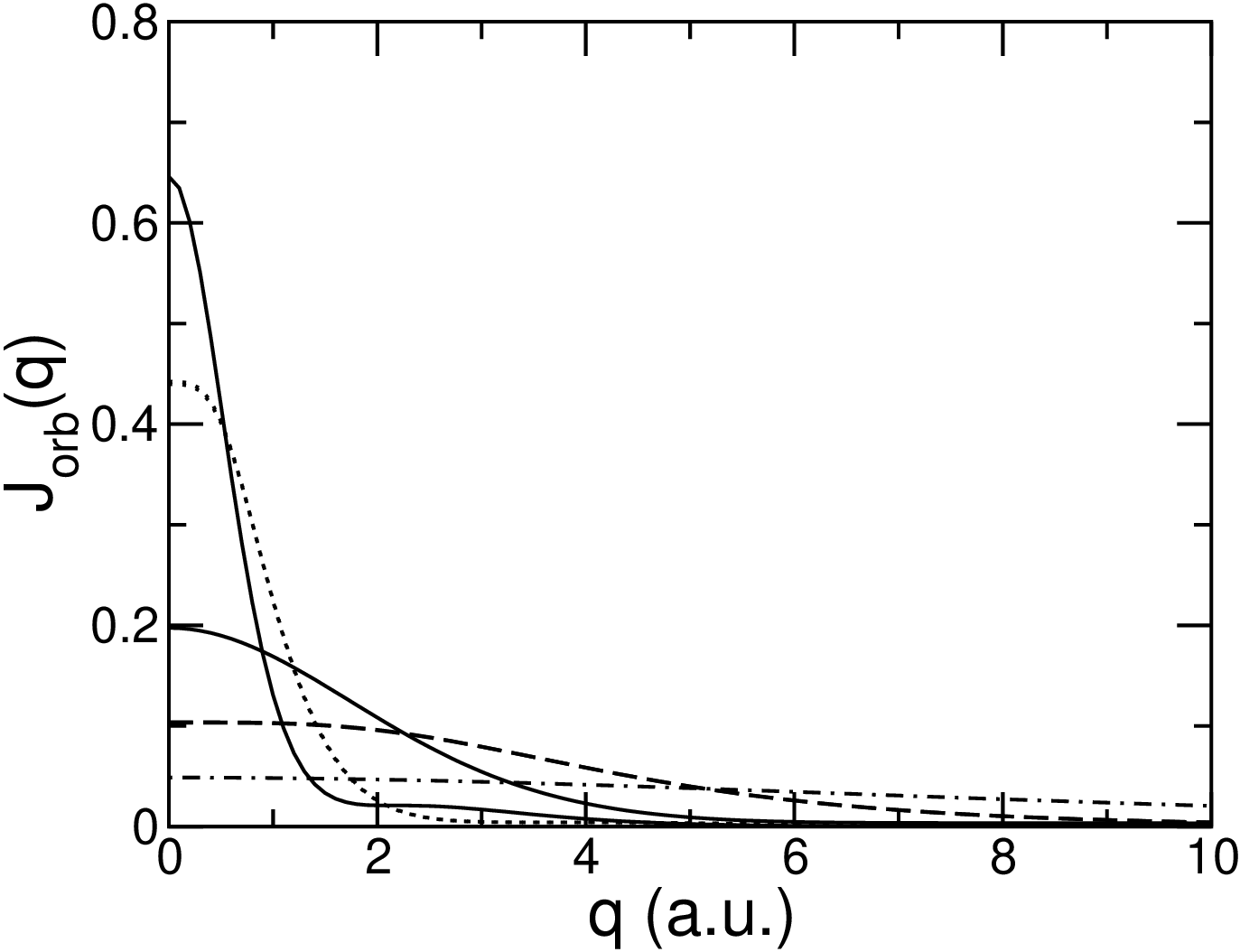}

\caption{Orbital Compton profiles of Ar, plotted as functions of the momentum
transfer $q$. In the decreasing order of the value of $J_{\mbox{orb}}(q=0)$,
the profiles correspond to $3s$, $3p_{3/2}/3p_{1/2}$, $2s$, $2p_{3/2}/2p_{1/2}$,
and $1s$ orbitals. Note that for all the cases, profiles of $p_{3/2}$
and $p_{1/2}$ orbitals are virtually identical. These profiles were
computed using the universal basis set\cite{basis-malli}.}

\label{fig-orb-ar}
\end{figure}

Additionally, in Figs. \ref{fig-ar} and \ref{fig-orb-ar}, respectively,
we present our total and orbital CPs of Ar plotted as a function of
the momentum transfer $q$. From Ar onwards, CP results of Mendelsohn
\emph{et al.} \cite{rel-com-prof} \emph{}exist\emph{,} which were
computed from the DHF orbitals obtained from finite-difference-based
calculations. If our calculated CPs are correct, they should be in
good agreements with those of Mendelsohn \emph{et al.}\cite{rel-com-prof}.
Therefore, it is indeed heartening for us to note that our CP results
computed with the universal basis set\cite{basis-malli} are in perfect
agreement with those of Mendelsohn \emph{et al.}\cite{rel-com-prof}
to the decimal places, and for the $q$ points, reported by them.
As a matter of fact even our CPs obtained using the smaller well-tempered
basis set\cite{basis-mat}, disagree with those of Mendelsohn \emph{et
al.}\cite{rel-com-prof} by very small amounts. Thus, this gives us
confidence about the essential correctness of our approach.

When compared to the experiments, for $q=0$, our value of CP of $5.054$
computed with universal basis set, is in excellent agreement with
the experimental value of $5.058$\cite{cp-exp-1}. For $0.1$a.u.$\leq q\leq$$0.8$a.u.
our results begin to overestimate the experimental ones slightly.
For $q\geq$$0.9$a.u., however, our theoretical results underestimate
the experimental results by small amounts. The nonrelativistic HF
results\cite{he-hf} also exhibit the same pattern with respect to
the experimental results. Upon comparing our CPs to the nonrelativistic
HF CPs\cite{he-hf}, we notice that the two sets of values differ
slightly for smaller values of $q$. However, the difference between
the two begins to become insignificant as we approach larger values
of $q$, suggesting that the relativistic effects will be most prominent
for $q\approx0$.

Finally we examine the contributions of the individual orbitals to
the atomic CP in Fig. \ref{fig-orb-ar}, which presents the orbital
profiles of all the orbitals of Ar. We observe the following trends:
(i) $3s$ profile has the maximum value at $q=0$, followed by $3p_{3/2}/3p_{1/2}$
profiles. The minimum value at $q=0$ corresponds to the $1s$ profile.
(ii) Profiles of outer orbitals vary more rapidly with $q$, as compared
to the inner ones. In other words, profile flattening occurs as one
moves inwards from the valence to the core orbitals. (iii) Again no
significant fine-structure splitting is observed, in that the profiles
of $np_{3/2}$ and $np_{1/2}$ orbitals differed from each other by
small amounts, pointing to the smallness of relativistic effects.

\subsection{Kr}

Now, we discuss our DHF results of Compton profile of Kr. The DHF
calculations on Kr atom were performed using the following two basis
sets: (i) smaller $(20s$,$15p$, $9d$) basis set of Koga \emph{et
al.}\cite{tate-1}, and the (ii) large ($32s$, $29p,$ $20d$) universal
basis set of Malli \emph{et al.}\cite{basis-malli}. Calculated total
CPs of Kr, for $0$ a.u.$\leq q\leq30$ a.u., are presented in table
\ref{tab-kr}, which also contains the nonrelativistic HF profiles
computed by Clementi and Roetti\cite{he-hf}, DHF profiles calculated
by Mendelsohn \emph{et al.}\cite{rel-com-prof}, and the experimental
results of Eisenberger and Reed\cite{cp-exp-1}.

\begin{table}

\caption{Our results on total profiles of Kr computed using the smaller basis
set of Koga, Tatewaki and Matsuoka (KTM)\cite{tate-1}, and the universal
basis set\cite{basis-malli}. Relativistic results of other authors\cite{rel-com-prof},
nonrelativistic HF results\cite{he-hf}, and the experimental results\cite{cp-exp-1}
are also presented for comparison.}

\begin{tabular}{|c|c|c|c|c|c|}
\hline 
$q$(a.u.)&
$J(q)$(KTM)$^{a}$&
$J(q)$(Uni)$^{b}$&
$J(q)$(DHF)$^{c}$&
$J(q)$(HF)$^{d}$&
$J(q)$(Exp.)$^{e}$\tabularnewline
\hline
\hline 
$0.0$&
$7.1788$&
$7.1871$&
$7.19$&
$7.228$&
$7.205$\tabularnewline
\hline 
$0.1$&
$7.1470.$&
$7.1548$&
$7.15$&
$7.194$&
$7.152$\tabularnewline
\hline 
$0.2$&
$7.0452$&
$7.0505$&
$7.05$&
$7.085$&
$7.022$\tabularnewline
\hline 
$0.3$&
$6.8588$&
$6.8595$&
---&
$6.888$&
$6.767$\tabularnewline
\hline 
$0.4$&
$6.5780$&
$6.5735$&
$6.57$&
$6.595$&
$6.459$\tabularnewline
\hline 
$0.5$&
$6.2087$&
$6.2010$&
---&
$6.216$&
$6.098$\tabularnewline
\hline 
$0.6$&
$5.7744$&
$5.7670$&
$5.77$&
$5.776$&
$5.701$\tabularnewline
\hline 
$0.7$&
$5.3093$&
$5.3053$&
---&
$5.309$&
$5.289$\tabularnewline
\hline 
$0.8$&
$4.8485$&
$4.8486$&
---&
$4.848$&
$4.880$\tabularnewline
\hline 
$0.9$&
$4.4197$&
$4.4225$&
---&
$4.420$&
$4.491$\tabularnewline
\hline 
$1.0$&
$4.0395$&
$4.0429$&
$4.04$&
$4.039$&
$4.133$\tabularnewline
\hline 
$1.2$&
$3.4425$&
$3.4432$&
---&
$3.441$&
$3.540$\tabularnewline
\hline 
$1.4$&
$3.0368$&
$3.0353$&
---&
$3.037$&
$3.122$\tabularnewline
\hline 
$1.6$&
$2.7662$&
$2.7650$&
---&
$2.769$&
$2.850$\tabularnewline
\hline 
$1.8$&
$2.5787$&
$2.5785$&
---&
$2.583$&
$2.670$\tabularnewline
\hline 
$2.0$&
$2.4362$&
$2.4364$&
$2.44$&
$2.441$&
$2.533$\tabularnewline
\hline 
$2.5$&
$2.1425$&
$2.1428$&
---&
$2.144$&
$2.219$\tabularnewline
\hline 
$3.0$&
$1.8571$&
$1.8572$&
---&
$1.857$&
$1.898$\tabularnewline
\hline 
$3.5$&
$1.5784$&
$1.5782$&
---&
$1.578$&
$1.597$\tabularnewline
\hline
$4.0$&
$1.3257$&
$1.3255$&
$1.33$&
$1.326$&
$1.338$\tabularnewline
\hline
$5.0$&
$0.9333$&
$0.9335$&
---&
$0.934$&
$0.937$\tabularnewline
\hline
$6.0$&
$0.6773$&
$0.6773$&
$0.677$&
$0.678$&
$0.683$\tabularnewline
\hline
$7.0$&
$0.5118$&
$0.5118$&
---&
$0.512$&
$0.522$\tabularnewline
\hline
$8.0$&
$0.4001$&
$0.4001$&
---&
$0.400$&
$0.399$\tabularnewline
\hline
$9.0$&
$0.3205$&
$0.3205$&
---&
$0.319$&
$0.316$\tabularnewline
\hline
$10.0$&
$0.2608$&
$0.2608$&
$0.261$&
$0.259$&
$0.254$\tabularnewline
\hline
$15.0$&
$0.1062$&
$0.1062$&
---&
$0.104$&
$0.095$\tabularnewline
\hline
$20.0$&
$0.0506$&
$0.0506$&
---&
$0.049$&
$0.044$\tabularnewline
\hline
$25.0$&
 $0.0271$&
$0.0271$&
$0.027$&
$0.026$&
$0.022$\tabularnewline
\hline
$30.0$&
$0.0157$&
$0.0157$&
---&
$0.015$&
$0.009$\tabularnewline
\hline
\end{tabular}

\begin{raggedright}
$^{a}$DHF results computed using the basis set of Koga, Tatewaki
and Matsuoka\cite{tate-1}.
\par\end{raggedright}

\begin{raggedright}
$^{b}$DHF results computed using the universal basis set\cite{basis-malli}
\par\end{raggedright}

\begin{raggedright}
$^{c}$DHF results of Mendelsohn \emph{et al.}\cite{rel-com-prof}
based upon finite-difference calculations
\par\end{raggedright}

\begin{raggedright}
$^{d}$Nonrelativistic HF results from Ref.\cite{he-hf}
\par\end{raggedright}

\begin{raggedright}
$^{e}$Experimental results from Ref.\cite{cp-exp-1}
\par\end{raggedright}

\label{tab-kr}
\end{table}

\begin{figure}
\includegraphics[width=10cm,keepaspectratio]{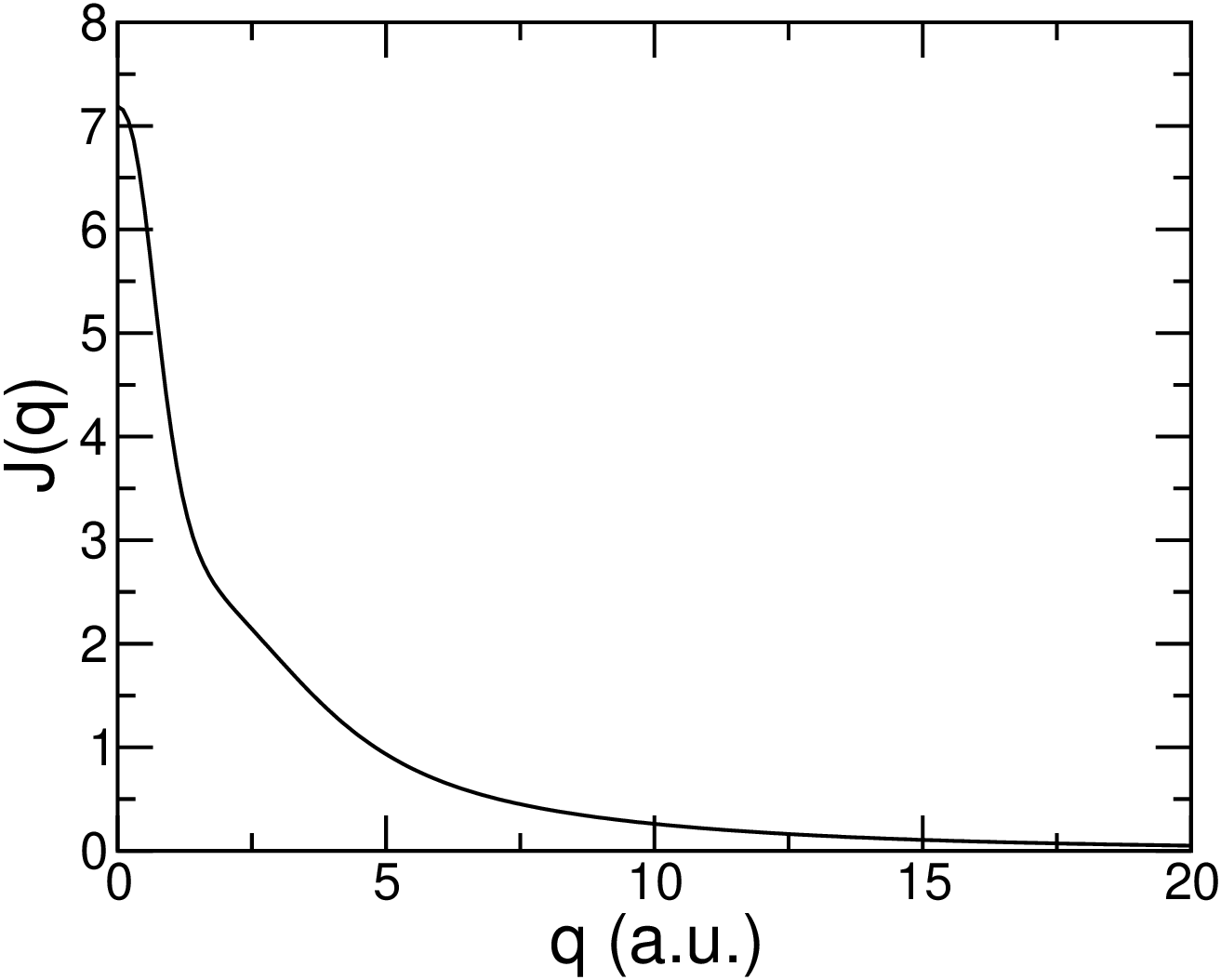}

\caption{DHF Compton profile of Kr, $J(q)$, computed using the universal
basis set\cite{basis-malli}, plotted as a function of the momentum
transfer $q$. }

\label{fig-kr}
\end{figure}

\begin{figure}
\includegraphics[width=10cm,keepaspectratio]{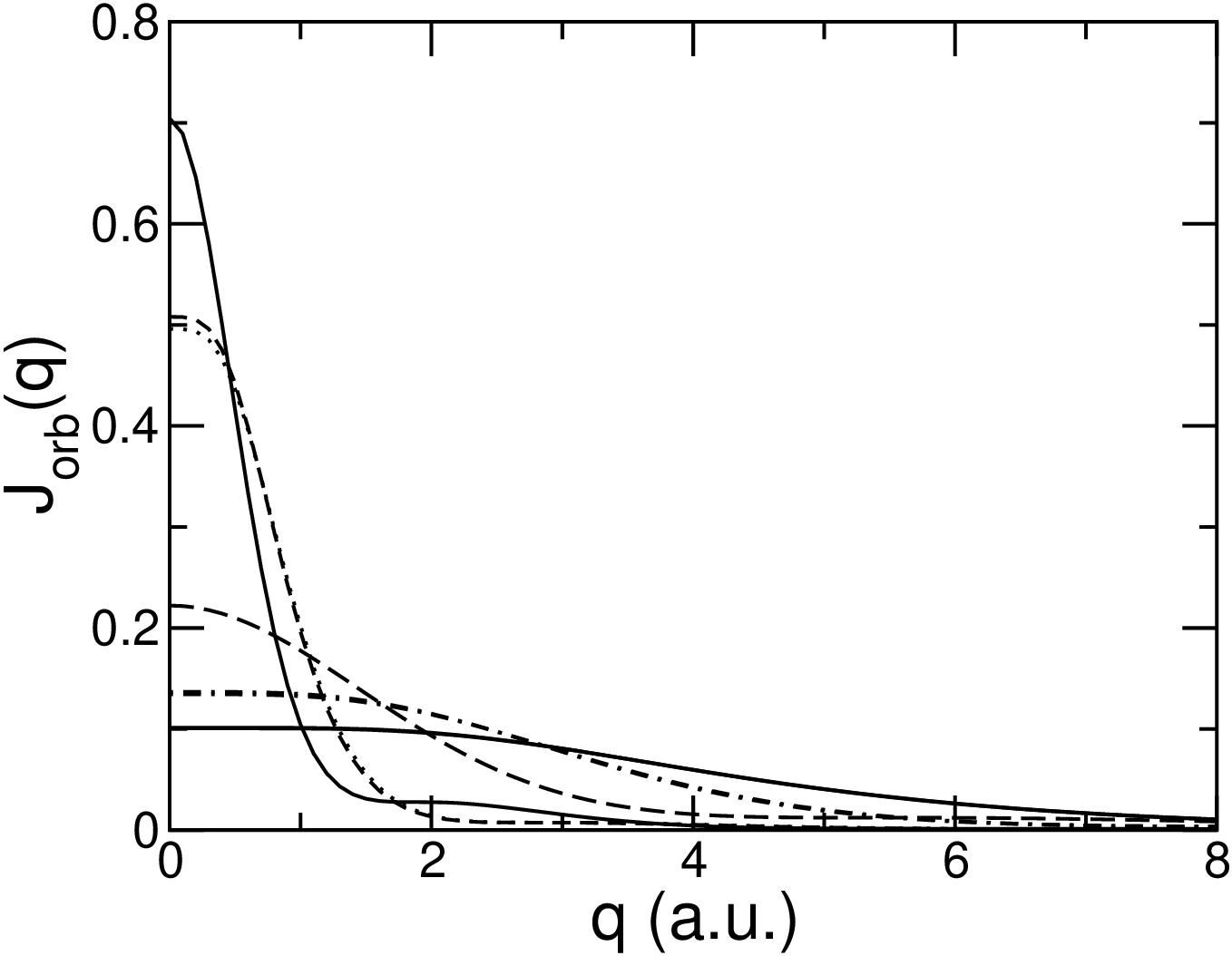}

\caption{Orbital Compton profiles of Kr for $4s$, $4p_{3/2}$, $4p_{1/2}$,
$3s$, $3p_{3/2}$/$3p_{1/2}$, and $3d_{5/2}$/$3d_{3/2}$ orbitals
in the order of decreasing values at $q=0$. For small $q$ values,
the differences between the $4p_{3/2}$ and $4p_{1/2}$ profiles are
visible. These profiles were computed using the universal basis set\cite{basis-malli}.}

\label{fig-orb-kr}
\end{figure}

In Figs. \ref{fig-kr} and \ref{fig-orb-kr}, respectively, our total
and orbital CPs of Kr, are plotted as a function of the momentum transfer
$q$. Upon comparing our CPs of Kr obtained using two basis sets we
note that: (i) for small values of $q$, the values obtained using
the smaller basis set of Koga \emph{et al.}\cite{tate-1} are slightly
smaller than the ones obtained using the universal basis set, and
(ii) for large values of $q$, the results obtained using the two
basis sets are in excellent agreement with each other. Next, we compare
our calculated CPs with those computed by Mendelsohn \emph{et al.}\cite{rel-com-prof}
using the numerical orbitals obtained in their DHF calculations. From
table \ref{tab-kr} it is obvious that, for the all the $q$ values
for which Mendelsohn \emph{et al.}\cite{rel-com-prof} reported their
CPs, our profiles obtained using the universal basis set\cite{basis-malli},
are in exact agreement with their results. As a matter of fact, the
agreement between the results of Mendelsohn \emph{et al.}\cite{rel-com-prof},
and our results computed using the smaller basis set of Koga \emph{et
al.}\cite{tate-1}, is also excellent.

Upon comparing our results to experimental ones, we see that our universal
basis set value of $J(q=0)=7.187$, is in excellent agreement with
the experimental value of $7.205$\cite{cp-exp-1}. For other values
of momentum transfer in the range $0.1$ a.u.$\leq q\leq$$1.0$ a.u.,
although the agreement between our results and the experiments is
slightly worse, yet our results are closer to the experimental value
as compared to the nonrelativistic HF results\cite{he-hf}. For higher
values of momentum transfer, our DHF results are fairly close to the
HF results suggesting that in the region of large $q$, relativistic
effects are unimportant. Thus, we conclude that from Kr onwards, relativistic
effects make their presence felt in the small $q$ region.

Finally, we investigate the orbital CPs of Kr in Fig. \ref{fig-orb-kr},
which presents the plots of the profiles of outer orbitals starting
from $3d_{3/2}$ to $4p_{3/2}$. As far as the general trends of the
orbital profiles are concerned, they are similar to what we observed
for the cases of Ne and Ar, except for one important aspect. Unlike
the Ne and Ar, for Kr for the first time we begin to observe the fine
structure splitting in the orbital profiles of $4p_{3/2}$ and $4p_{1/2}$
orbitals in the low $q$ region, as is obvious from Fig. \ref{fig-orb-kr}.
For example, for $q=0$, corresponding values are $J_{4p_{3/2}}=0.508$,
and $J_{4p_{1/2}}=0.496$, amounting to a difference of $\approx2\%$.
This is in complete agreement with our earlier observation that the
relativistic effects make significant contributions to the CPs of
Kr in the small $q$ region.

\subsection{Xe }

\label{sub-sec:xe}

In this section, we discuss our results on the relativistic Compton
profiles of Xe. The DHF calculations on Xe atom were performed using
the following two basis sets: (i) smaller $(22s$,$18p$, $12d$)
basis set of Koga \emph{et al.}\cite{tate-1}, and the (ii) large
($32s$, $29p,$ $20d$) universal basis set of Malli \emph{et al.}\cite{basis-malli}.
Total CPs of Xe, for selected values of momentum transfer in the range
$0$ a.u.$\leq q\leq100$ a.u., are presented in table \ref{tab-xe}.
For the sake of comparison, the same table also contains DHF, and
the nonrelativistic HF, profiles calculated by Mendelsohn \emph{et
al.}\cite{rel-com-prof}. Here, we are unable to compare our results
with the experiments, because, to the best of our knowledge, no experimental
measurements of the CPs of Xe exist. 

\begin{table}

\caption{Total CPs of Xe computed using the smaller basis set of Koga, Tatewaki
and Matsuoka (KTM)\cite{tate-1}, and the universal basis set\cite{basis-malli}.
Relativistic results of other authors\cite{rel-com-prof}, and nonrelativistic
HF results\cite{rel-com-prof} are also presented for comparison.}

\begin{tabular}{|c|c|c|c|c|}
\hline 
$q$(a.u.)&
$J(q)$(KTM)$^{a}$&
$J(q)$(Uni)$^{b}$&
$J(q)$(DHF)$^{c}$&
$J(q)$(HF)$^{d}$\tabularnewline
\hline
\hline 
$0.0$&
$9.722$&
$9.737$&
$9.74$&
$9.88$\tabularnewline
\hline 
$0.1$&
$9.673$&
$6.687$&
$9.69$&
$9.82$\tabularnewline
\hline 
$0.2$&
$9.515$&
$9.523$&
$9.52$&
$9.65$\tabularnewline
\hline 
$0.4$&
$8.784$&
$8.775$&
$8.78$&
$8.85$\tabularnewline
\hline 
$0.6$&
$7.597$&
$7.587$&
$7.59$&
$7.62$\tabularnewline
\hline 
$1.0$&
$5.448$&
$5.451$&
$5.45$&
$5.46$\tabularnewline
\hline 
$1.5$&
$4.293$&
$4.292$&
$4.29$&
$4.31$\tabularnewline
\hline 
$2.0$&
$3.678$&
$3.678$&
$3.68$&
$3.69$\tabularnewline
\hline 
$4.0$&
$1.707$&
$1.707$&
$1.71$&
$1.72$\tabularnewline
\hline
$6.0$&
$1.060$&
$1.061$&
$1.06$&
$1.06$\tabularnewline
\hline 
$10.0$&
$0.5150$&
$0.5150$&
$0.515$&
$0.515$\tabularnewline
\hline 
$25.0$&
$0.0660$&
$0.0662$&
$0.066$&
$0.064$\tabularnewline
\hline 
$50.0$&
$0.0088$&
$0.0088$&
$0.0088$&
$0.0076$\tabularnewline
\hline 
$100.0$&
$0.00067$&
$0.0067$&
$0.00068$&
$0.00043$\tabularnewline
\hline
\end{tabular} 

\begin{raggedright}
$^{a}$our DHF results computed using the basis set of Koga, Tatewaki
and Matsuoka\cite{tate-1}.
\par\end{raggedright}

\begin{raggedright}
$^{b}$our DHF results computed using the universal basis set\cite{basis-malli}
\par\end{raggedright}

\begin{raggedright}
$^{c}$DHF results of Mendelsohn \emph{et al.}\cite{rel-com-prof}
based upon finite-difference calculations
\par\end{raggedright}

\begin{raggedright}
$^{d}$Nonrelativistic HF results reported in Ref.\cite{rel-com-prof}
\par\end{raggedright}

\label{tab-xe}
\end{table}

\begin{figure}
\includegraphics[width=10cm,keepaspectratio]{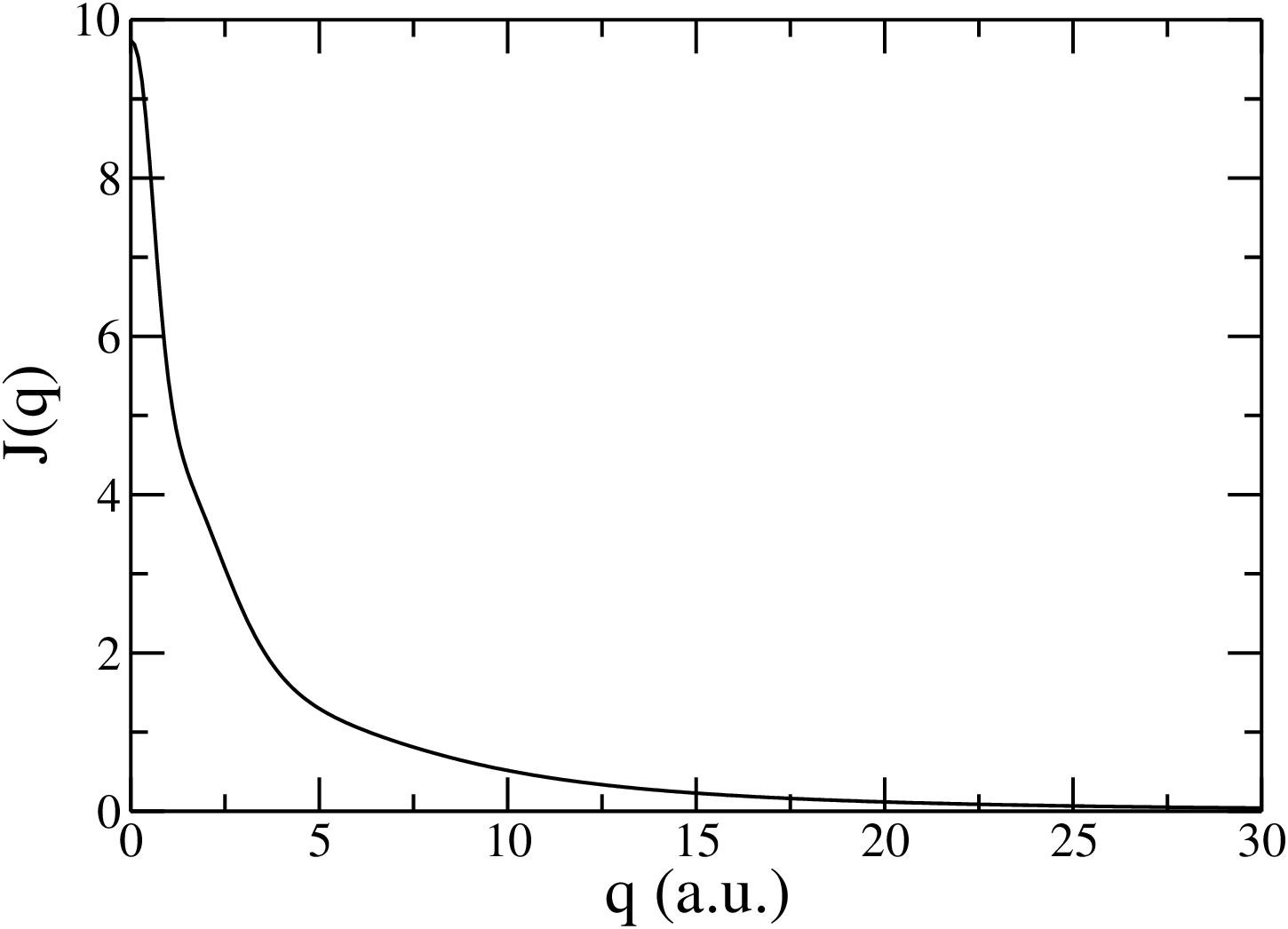}

\caption{DHF Compton profile of Xe, computed using the universal basis set\cite{basis-malli},
and plotted as a function of the momentum transfer $q$.}

\label{fig-xe}
\end{figure}

\begin{figure}
\includegraphics[width=10cm,keepaspectratio]{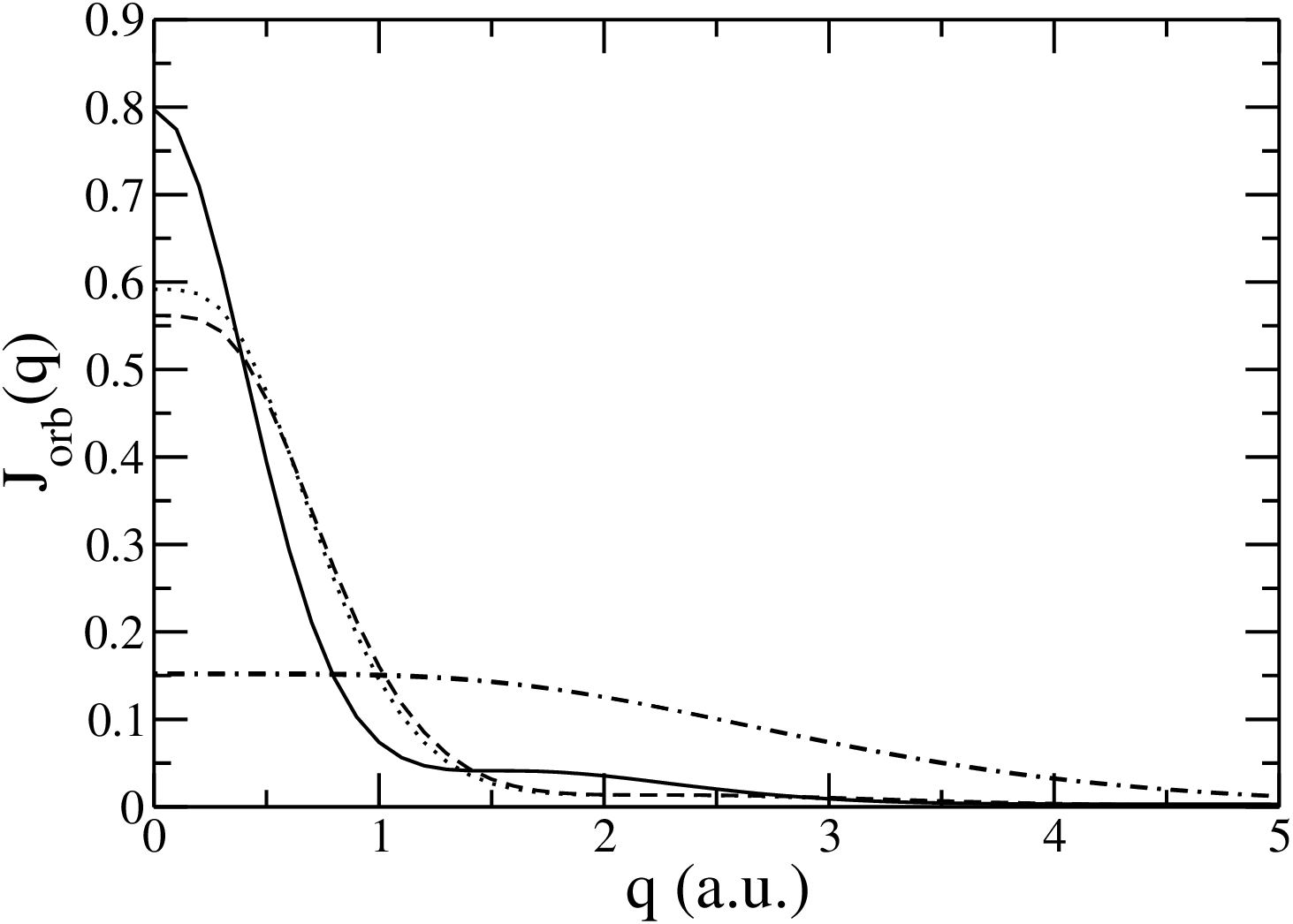}

\caption{Orbital Compton profiles of Xe for $5s$, $5p_{3/2}$, $5p_{1/2}$,
and $4d_{5/2}$/$4d_{3/2}$ orbitals in the order of decreasing values
at $q=0$. For small $q$ values, the differences between the $5p_{3/2}$
and $5p_{1/2}$ profiles are quite significant. These profiles were
computed using the universal basis set\cite{basis-malli}}

\label{fig-orb-xe}
\end{figure}

Additionally, in Figs. \ref{fig-xe} and \ref{fig-orb-xe}, respectively,
we present the plots of our total and orbital CPs of Xe. Upon comparing
our total CPs obtained using the two basis sets we find that, as before,
they disagree for smaller values of $q$, with the CPs obtained using
the smaller basis set\cite{tate-1} being slightly lower than those
obtained using the universal basis set\cite{basis-malli}. As is obvious
from table \ref{tab-xe}, that for $q\geq1.5$ a.u., the two sets
of basis functions yield virtually identical results. In the same
table, when we compare our results to the earlier DHF results of Mendelsohn
\emph{et al.}\cite{rel-com-prof}, we find that for all the $q$ values,
the agreement between our universal basis-set based CPs, and their
results, is perfect up to the decimal places reported by them. This
again points to the correctness of our calculations. 

Upon comparing our DHF results to the nonrelativistic HF results of
Mendelsohn \emph{et al.}\cite{rel-com-prof}, we find that for smaller
values of $q$, the DHF values of CPs are smaller than the HF values,
while for large values of $q$, the trend is just the opposite. 

Finally, upon examining the orbital profiles presented in Fig. \ref{fig-orb-xe},
we observe further evidence of the importance of relativistic effects
in Xe. As is obvious from the figure, the fine-structure splitting
between the orbitals profiles of $5p_{3/2}$ and $5p_{1/2}$ orbitals
is larger as compared to $4p_{3/2}/4p_{1/2}$ splitting in Kr, and
persists for a longer range of $q$ values. For smaller values of
$q$, $J_{5p_{3/2}}(q)>J_{5p_{1/2}}(q)$, while for large $q$ values,
opposite is the case. For $q=0,$ $J_{5p_{3/2}}=0.592$, while $J_{5p_{1/2}}=0.562$,
which amounts to a difference of $\approx5\%$.

\subsection{Rn}

\label{subsec-Rn}

As far as atomic Rn is concerned, to the best of our knowledge, no
prior experimental studies of its Compton profiles exist. However,
Biggs \emph{et al.}\cite{rel-comp-prof-2}did perform DHF calculations
of this atom, using a finite difference approach, with which we compare
our results later on in this section. Our DHF calculations on Rn atom
were performed using the following two basis sets: (i) smaller $(25s$,$21p$,
$15d$, $10f$) basis set of Koga \emph{et al.}\cite{tate-2}, and
the (ii) large ($32s$, $29p,$ $20d$, $15f$) universal basis set
of Malli \emph{et al.}\cite{basis-malli}. Total CPs of Rn, for selected
values of momentum transfer in the range $0$ a.u.$\leq q\leq100$
a.u., are presented in table \ref{tab-rn}. 

\begin{table}

\caption{Total CPs of Rn computed using the smaller basis set of Koga, Tatewaki
and Matsuoka (KTM)\cite{tate-2}, and the universal basis set\cite{basis-malli},
compared to the earlier calculations of Biggs \emph{et al.} \cite{rel-comp-prof-2}.}

\begin{tabular}{|c|c|c|c|}
\hline 
$q$(a.u.)&
$J(q)$(KTM)$^{a}$&
$J(q)$(Uni)$^{b}$&
$J(q)$(DHF)$^{c}$\tabularnewline
\hline
\hline 
$0.0$&
$11.8344$&
$11.8531$&
$11.9$\tabularnewline
\hline 
$0.1$&
$11.7850$&
$11.8026$&
$11.8$\tabularnewline
\hline 
$0.2$&
$11.6176$&
$11.6306$&
$11.6$\tabularnewline
\hline 
$0.4$&
$10.8055$&
$10.7996$&
$10.8$\tabularnewline
\hline 
$0.6$&
$9.4877$&
$9.4744$&
$9.47$\tabularnewline
\hline 
$1.0$&
$7.2130$&
$7.2130$&
$7.21$\tabularnewline
\hline 
$1.6$&
$5.8127$&
$5.8132$&
$5.81$\tabularnewline
\hline 
$2.0$&
$5.1530$&
$5.1533$&
$5.15$\tabularnewline
\hline 
$4.0$&
$2.8379$&
$2.8381$&
$2.84$\tabularnewline
\hline 
$6.0$&
$2.0453$&
$2.0454$&
$2.05$\tabularnewline
\hline 
$10.0$&
$0.9804$&
$0.9804$&
$0.98$\tabularnewline
\hline 
$30.0$&
$0.1083$&
$0.1083$&
$0.11$\tabularnewline
\hline 
$60.0$&
$0.0166$&
$0.0166$&
$0.017$\tabularnewline
\hline 
$100.0$&
$0.0037$&
$0.0037$&
$0.0037$\tabularnewline
\hline
\end{tabular}

\label{tab-rn}

\begin{raggedright}
$^{a}$our DHF results computed using the basis set of Koga, Tatewaki
and Matsuoka\cite{tate-2}.
\par\end{raggedright}

\begin{raggedright}
$^{b}$our DHF results computed using the universal basis set\cite{basis-malli}
\par\end{raggedright}

\begin{raggedright}
$^{c}$DHF results of Biggs \emph{et al.}\cite{rel-comp-prof-2} based
upon finite-difference calculations
\par\end{raggedright}
\end{table}

\begin{figure}
\includegraphics[width=10cm,keepaspectratio]{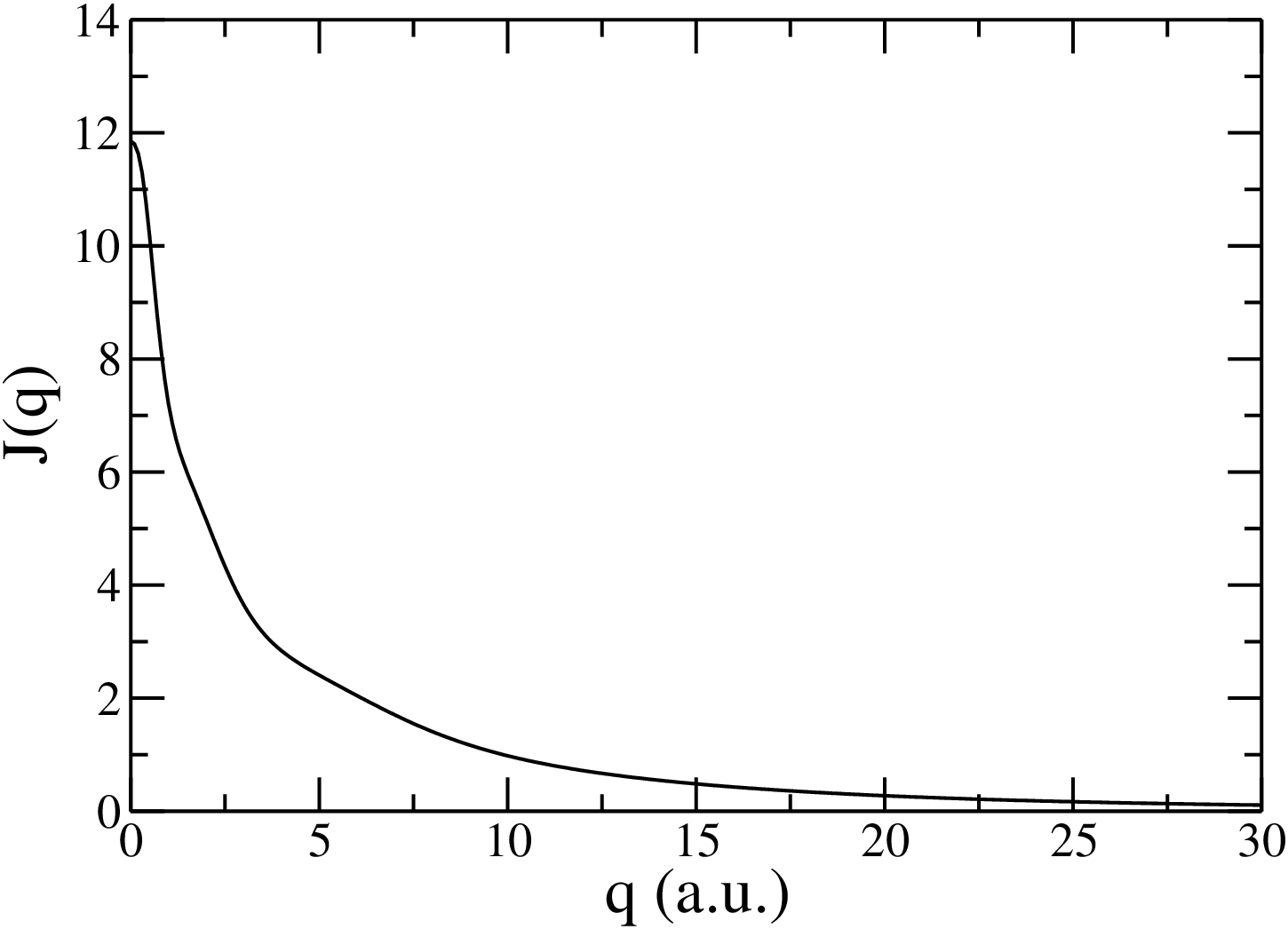}

\caption{DHF Compton profile of Rn, computed using the universal basis set\cite{basis-malli},
and plotted as a function of the momentum transfer $q$.}

\label{fig-rn}
\end{figure}

\begin{figure}
\includegraphics[width=10cm,keepaspectratio]{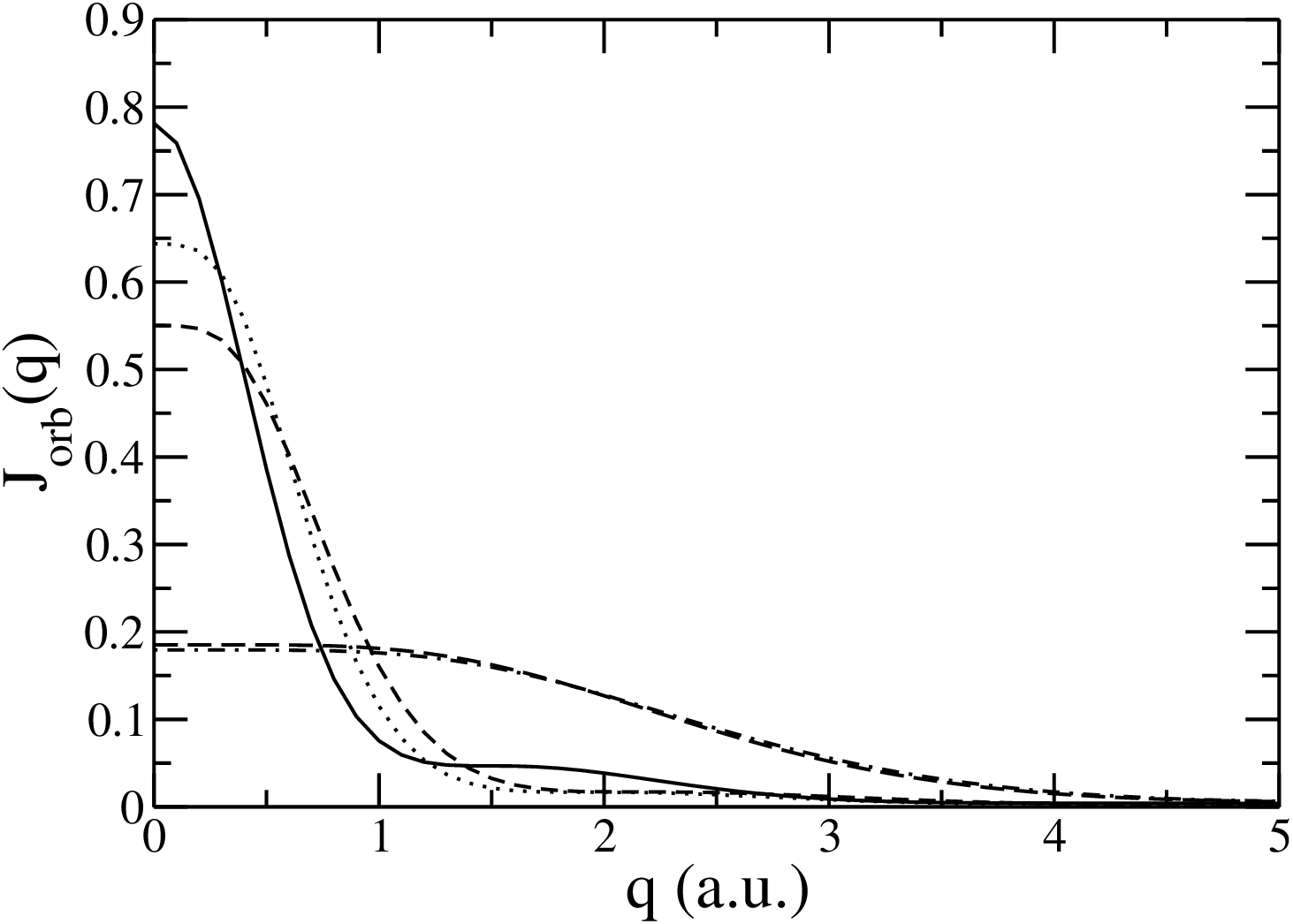}

\caption{Orbital Compton profiles of Rn for $6s$, $6p_{3/2}$, $6p_{1/2}$,
and $5d_{5/2}$/$5d_{3/2}$ orbitals in the order of decreasing values
at $q=0$. For small $q$ values, the differences between the $6p_{3/2}$
and $6p_{1/2}$ profiles are quite large. Even the splitting of $5d_{5/2}$
and $5d_{3/2}$ profiles is visible. These profiles were computed
using the universal basis set\cite{basis-malli}.}

\label{fig-orb-rn}
\end{figure}

Our results for total and orbital CPs of Rn are plotted in Figs. \ref{fig-rn}
and \ref{fig-orb-rn}, respectively. As for other atoms, we find that
our total CPs obtained using the two basis sets disagree for smaller
values of $q$, with the CPs obtained using the smaller basis set
of Koga \emph{et al.}\cite{tate-2} being slightly smaller than those
obtained using the universal basis set\cite{basis-malli}. From table
\ref{tab-rn} we deduce that for $q\geq4.0$ a.u., the two sets of
basis functions yield virtually identical values of CPs. In the same
table, when we compare our results to the earlier DHF calculations
of Biggs \emph{et al.}\cite{rel-comp-prof-2}, we find that for all
the $q$ values, the agreement between our universal basis-set based
CPs, and their results, is perfect up to the decimal places reported
by them. 

Of all the rare gas atoms considered so far, on the intuitive grounds
we expect the relativistic effects to be the strongest in Rn. Indeed,
this is what we confirm upon investigating the orbital profiles presented
in Fig. \ref{fig-orb-rn}. As is obvious from the figure, the splitting
between the orbitals profiles of $6p_{3/2}$ and $6p_{1/2}$ orbitals
is quite big, and persists for a large range of $q$ values. Similar
to the case of Xe, here also for smaller values of $q$, $J_{6p_{3/2}}(q)>J_{6p_{1/2}}(q)$,
while for large $q$ values, opposite is the case. For $q=0,$ $J_{6p_{3/2}}=0.644$,
while $J_{6p_{1/2}}=0.551$, amounting to a difference of $\approx15\%$,
which is quite substantial. The fine-structure splitting between the
profiles of $5d_{5/2}$ and $5d_{3/2}$ orbitals although is not quite
that large, yet it is visible in Fig. \ref{fig-orb-rn}. At $q=0,$$J_{5d_{5/2}}=0.185$,
and $J_{5d_{3/2}}=0.179$, leading to a difference of $\approx3\%$,
which is quite significant for an inner orbital. Thus, we conclude
that the relativistic effects are quite substantial in case of Rn,
and, therefore, it will be useful if experiments are performed on
this system to ascertain this.

\subsection{Z dependence of relativistic effects on Compton Profiles}

\begin{figure}
\includegraphics[width=10cm,keepaspectratio]{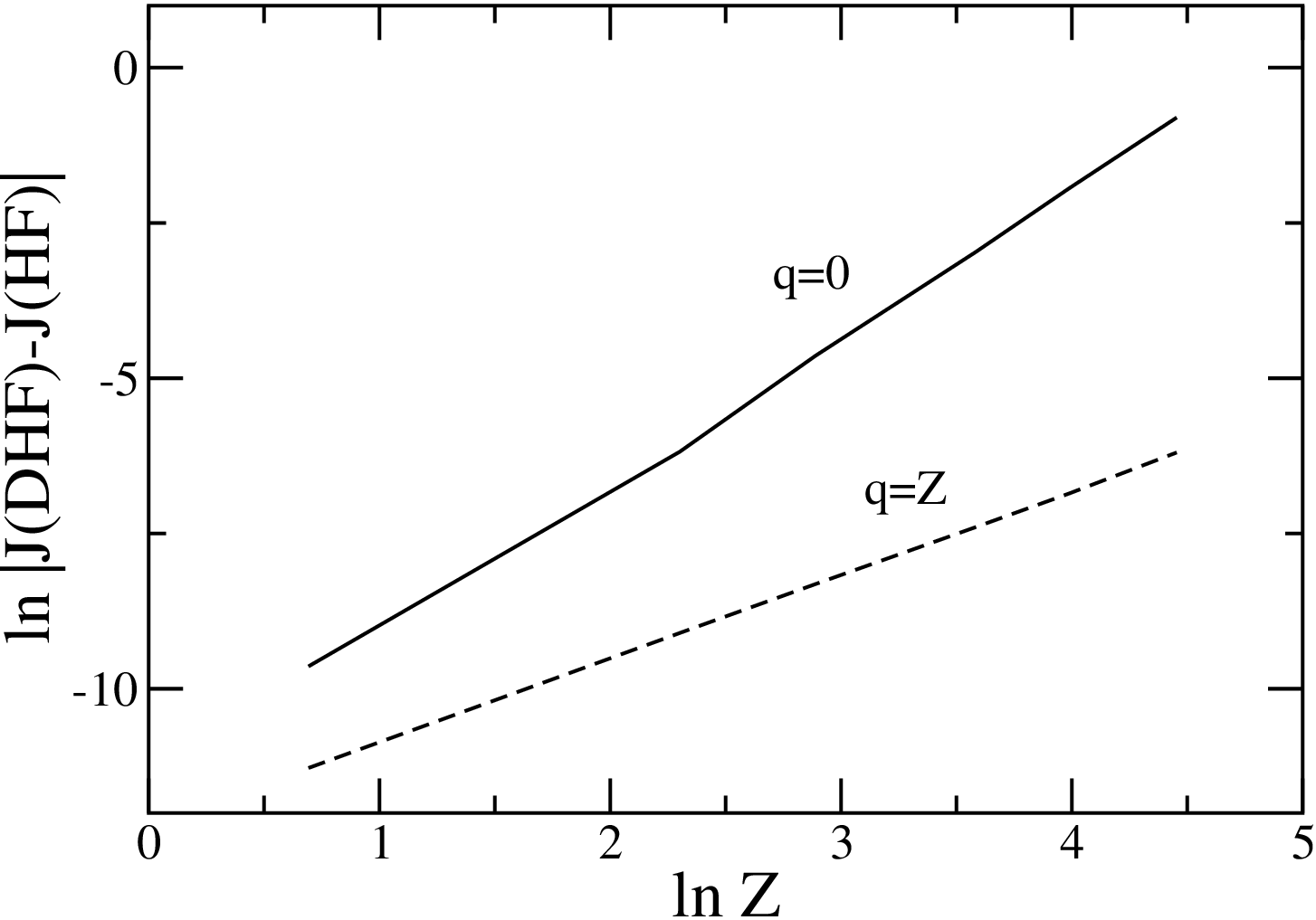}\label{fig-logz}

\caption{Difference between the relativistic ($J(\mbox{DHF})$), and the nonrelativistic
($J(\mbox{HF})$) Compton profiles plotted, on a logarithmic scale,
as a function of the atomic number $Z$. Plots correspond to the momentum
transfer values $q=0$, and $q=Z$.}
\end{figure}

In earlier sections, while discussing relativistic effects on Compton
profiles, we noticed that they were most prominent for small momentum
transfers. Moreover, one intuitively expects the relativistic effects
to increase with increasing atomic number $Z$. In this section our
aim is to perform a quantitative investigation of relativistic effects
on quantum profiles, as a function of $Z$, for both large and small
values of momentum transfer. We noticed that for small momentum transfers,
DHF values of $J$ were smaller than their nonrelativistic counterparts,
while for large momentum transfer opposite was the case. Therefore,
for a given value of momentum transfer $q$, we quantify relativistic
effects in terms of $|J(\mbox{DHF})-J(\mbox{HF})|$, which is the
magnitude of the difference of relativistic DHF value of the Compton
profile ($J(\mbox{DHF})$), and the nonrelativistic HF value of the
profile ($J(\mbox{HF})$). We obtain $J(\mbox{HF})$  by using a large
value of the velocity of light ($c=10^{4}$ a.u.) in the DHF calculations.
We explore the dependence of this quantity on $Z$, for two values
of momentum transfer, $q=0$, and $q=Z$ a.u., where the latter value
clearly belongs to the large momentum transfer regime. The values
of $\ln|J(\mbox{DHF})-J(\mbox{HF})|$ as a function of $\ln Z$, are
presented in Fig. \ref{fig-logz} for both these values of momentum
transfer. From the figure it obvious that, to a very good approximation,
the corresponding curves are straight lines, suggesting a power-law
dependence of the relativistic effects on $Z$. The slopes of the
least-square fit line for $q=0$ is $2.36$ while for $q=Z$, the
slope is $1.35$. Of course, these results are based upon data points
generated by six values of $Z$ (rare gas series), and consequently
can only be treated as suggestive. But the results suggest: (i) super-linear
dependence of the relativistic effect on quantum profiles in both
momentum transfer regimes, and (ii) stronger influence of relativity
in the small momentum transfer regime as compared to the large one.
Of course, this exploration can be refined further by separately investigating
the $Z$ dependence of these effects on the core and valence profiles.
Additionally, this investigation can be extended to a larger number
of atoms to obtain a larger set of data points. However, these calculations
are beyond the scope of the present work, and will be presented elsewhere. 

\section{Conclusions and Future Directions}

\label{sec-conclusions}

In this paper, we presented an approach aimed at computing the relativistic
Compton profiles of atoms within the DHF approximation, when the atomic
orbitals are represented as linear combinations of kinetically-balanced
set of Gaussian functions. The approach was applied to compute the
CPs of rare gas atoms ranging from He to Rn, and results were compared
to the experimental profiles, and theoretical profiles of other authors,
wherever such data was available. Additionally, the influence of size
and type of basis set was examined by performing calculations on each
atom with two basis sets: (i) a well-known smaller basis set, and
(ii) a large universal basis set proposed by Malli \emph{et al.}\cite{basis-malli}.

Upon comparing our results with the experiments, we found that for
lighter atoms He, Ne, and Ar, the agreement was similar to what one
obtains from the nonrelativistic HF calculations, indicating lack
of any significant relativistic effects for these atoms. For Kr, we
noticed that for smaller momentum transfer values, DHF results were
in better agreement with the experiments, as compared to the HF results.
For heavier atoms, Xe and Rn, unfortunately no experimental data is
available. Yet another quantitative indicator of the importance of
relativistic effects is the fine-structure splitting of the profiles,
\emph{i.e.}, the difference in the profiles of $np_{1/2}/np_{3/2}$
etc., which will have identical profiles in nonrelativistic calculations.
We found that this splitting becomes larger with the increasing atomic
number of the atom, thus justifying a relativistic treatment of the
problem for heavy atoms. Additionally, by comparing our results with
the nonrelativistic HF results we found that the relativistic effects
are most prominent in the region of small momentum transfer, while
at large momentum transfer, their contribution is much smaller.

In the literature, we were able to locate prior theoretical calculation
of relativistic CPs of atoms only from one group, namely the DHF calculations
of Mendelsohn \emph{et al.}\cite{rel-com-prof} and Biggs \emph{et
al.}\cite{rel-comp-prof-2}, performed on Ar, Kr, Xe, and Rn, employing
a finite-difference based approach. The CPs computed by \emph{}them\cite{rel-com-prof,rel-comp-prof-2}
for these atoms were found to be in perfect agreement with our results
computed using the universal basis set. This testifies to the correctness
of our approach, and suggests that by using a large basis set, it
is possible to reach the accuracy of finite-difference approaches
in relativistic calculations, not just on total energies\cite{basis-malli},
but also on expectation values. 

Having investigated the influence of the relativistic effects, the
next logical step will be to go beyond the mean-field DHF treatment,
and incorporate the influence of electron correlations on atomic CPs,
within a relativistic framework. Such a treatment can be within a
relativistic CI framework\cite{ci-shukla}, or can also be performed
within a perturbation-theoretic formalism. Work along these lines
is currently underway in our group, and the results will be submitted
for publication in future.

\appendix

\section{A Derivation of compton profile Matrix Elements over Kinetically
Balanced Gaussian Basis Sets}

\label{sec-appendix}

During our discussion here, we use the same notations for various
quantities as adopted in section \ref{sec-theory}.Our aim here is
to evaluate the closed form expressions for the following two integrals

\begin{equation}
J_{ij}^{L;\kappa}(q)=\frac{1}{2}\int_{q}^{\infty}pg_{\kappa i}^{L}(p)g_{\kappa j}^{L}(p)\, dp\label{jl}\end{equation}

\begin{equation}
J_{ij}^{S;\kappa}(q)=\frac{1}{2}\int_{q}^{\infty}pg_{\kappa i}^{S}(p)g_{\kappa j}^{S}(p)\, dp\label{js}\end{equation}

which, as explained in section \ref{sec-theory}, are needed to compute
the orbital (and total) atomic CPs when the KBGF based numerical formalism
is employed to solve the DHF equations. First, we will obtain expressions
for $g_{\kappa i}^{L}(p)$ and $g_{\kappa i}^{S}(p)$, the radial
Fourier transforms of the large and small component basis functions
$g_{\kappa i}^{L}(r)$ and $g_{\kappa i}^{S}(r)$, respectively, defined
as

\begin{equation}
g_{\kappa i}^{L}(p)=\frac{4\pi}{(2\pi)^{3/2}}\int_{0}^{\infty}rg_{\kappa i}^{L}(r)j_{l_{A}}(pr)\, dr\label{gl}\end{equation}

\begin{equation}
g_{\kappa i}^{S}(p)=\frac{4\pi}{(2\pi)^{3/2}}\int_{0}^{\infty}rg_{\kappa i}^{S}(r)j_{l_{B}}(pr)\, dr\label{gs}\end{equation}

where $j_{l_{A}}(pr)/j_{l_{A}}(pr)$ refer to the spherical Bessel
functions corresponding to the orbital angular momentum $l_{A}/l_{B}$
of the large/small component. The spherical Bessel function is related
to the Bessel function by the well-known relation

\begin{equation}
j_{\nu}(x)=\sqrt{\frac{\pi}{2x}}J_{\nu+1/2}(x),\label{bessel}\end{equation}

where $J_{\nu}(x)$ is the Bessel function.

\subsection{Derivation for the Large Component}

First , we obtain and expression for $g_{\kappa i}^{L}(p)$ by performing
the integral involved in Eq. (\ref{gl}). Substituting the expression
for $g_{\kappa i}^{L}(r)$ from Eq. (\ref{eq-glarge}) in Eq.(\ref{gl}),
we obtain 

\begin{eqnarray}
g_{\kappa i}^{L}(p) & = & \frac{4\pi}{(2\pi)^{3/2}}\int_{0}^{\infty}N_{\kappa i}^{L}r^{(n_{\kappa}+1)}e^{-\alpha_{i}r^{2}}j_{l_{A}}(pr)\, dr\nonumber \\
 & = & \frac{N_{\kappa i}^{L}}{\sqrt{p}}\int_{0}^{\infty}N_{\kappa i}^{L}r^{(n_{\kappa}+1/2)}e^{-\alpha_{i}r^{2}}J_{l_{A}+1/2}(pr)\, dr\label{*}\end{eqnarray}

where in the last step, we have used Eq. (\ref{bessel}). Next, on
using the relation $n_{\kappa}=l_{A}+1$, and the definite integral\cite{abram+steg}

\begin{equation}
\int_{0}^{\infty}x^{\nu+1}e^{-\alpha r^{2}}J_{\nu}(\beta x)\, dx=\frac{\beta^{\nu}}{(2\alpha)^{\nu+1}}e^{-\beta^{2}/4\alpha}[Re(\alpha)>0,Re(\nu)>0],\label{integral}\end{equation}

the Eq.(\ref{*}) simplifies to

\begin{equation}
g_{\kappa i}^{L}(p)=N_{\kappa i}^{L}\frac{p^{l_{A}}}{(2\alpha_{i})^{l_{A}+3/2}}e^{-p^{2}/4\alpha_{i}}.\end{equation}

On substituting the above result in Eq.(\ref{jl}), one obtains

\[
J_{ij}^{L;\kappa}(q)=\frac{1}{2}\int_{q}^{\infty}(N_{\kappa i}^{L})(N_{\kappa j}^{L})\frac{p^{2l_{A}+1}}{(4\alpha_{i}\alpha_{j})^{l_{A}+3/2}}e^{-p^{2}/4\alpha_{ij}}\, dp\]

where $\alpha_{ij}=\frac{\alpha_{i}\alpha_{j}}{\alpha_{i}+\alpha_{j}}$.
Next, on making the change of variable $t=\frac{p^{2}}{4\alpha_{ij}}$
in the integral above, leading to the lower limit $q_{t}=\frac{q^{2}}{4\alpha_{ij}}$,
we obtain\begin{eqnarray*}
J_{ij}^{L;\kappa}(q) & = & \frac{(N_{\kappa i}^{L})(N_{\kappa j}^{L})}{4}\frac{(4\alpha_{ij})^{l_{A}+1}}{(4\alpha_{i}\alpha_{j})^{l_{A}+3/2}}\int_{q_{t}}^{\infty}t^{l_{A}}e^{-t}\, dt,\end{eqnarray*}
leading to the final expression\begin{equation}
J_{ij}^{L;\kappa}(q)=\frac{(N_{\kappa i}^{L})(N_{\kappa j}^{L})}{4}\frac{(4\alpha_{ij})^{l_{A}+1}}{(4\alpha_{i}\alpha_{j})^{l_{A}+3/2}}\Gamma(l_{A}+1,q_{t}),\label{jl-final}\end{equation}

where $\Gamma(l_{A}+1,q_{t})$ is the incomplete gamma function. Since,
$l_{A}$ is a non-negative integer, the incomplete gamma function
can be easily computed using the series\cite{abram+steg},

\begin{equation}
\Gamma(l_{A}+1,q_{t})=(l_{A})!e^{-q_{t}}\sum_{m=0}^{l_{A}}\frac{{q_{t}}^{m}}{m!}.\label{igamma}\end{equation}

We note that our general result for $J_{ij}^{L;\kappa}(q)$ in Eq.
(\ref{jl-final}) leads to the same formulas as reported by Naon \emph{et
al.}\cite{naon} for the atomic CP matrix elements for s- and p-type
Gaussian orbitals, for the nonrelativistic case.

\subsection{Derivation for the small component}

Noting that the explicit form of the small component basis function
$g_{\kappa i}^{S}(r)$ (\emph{cf.} Eq. (\ref{eq-gsmall})) is\[
g_{\kappa i}^{S}(r)=N_{\kappa i}^{S}N_{\kappa i}^{L}\left[(n_{\kappa}+\kappa)r^{n_{\kappa}-1}e^{-\alpha_{i}r^{2}}-2\alpha_{i}r^{n_{\kappa}+1}e^{-\alpha_{i}r^{2}}\right].\]

On substituting the above in Eq.(\ref{gs}), the Fourier transform
of the small component basis function becomes\begin{equation}
g_{\kappa i}^{S}(p)=\frac{N_{\kappa i}^{S}N_{\kappa i}^{L}}{\sqrt{p}}\int_{0}^{\infty}\left[(n_{\kappa}+\kappa)r^{(n_{\kappa}-1/2)}-2\alpha_{i}r^{(n_{\kappa}+3/2)}\right]e^{-\alpha_{i}r^{2}}J_{l_{B}+1/2}(pr)\, dr\label{**}\end{equation}

As before, we seek a relation between $n_{\kappa}$ and $l_{B}$,
which is summarized in table \ref{tab-quantum}. Here, the two cases
have to be dealt separately since, the relations are different for
the two possibilities.

\begin{table}

\caption{Relationship between quantum numbers $\kappa$, $n_{\kappa}$, and
$l_{B}$, for relativistic atomic orbitals.}

\begin{tabular}{|c|c|c|}
\hline 
$\kappa$&
$n_{\kappa}$&
$l_{B}$\tabularnewline
\hline
\hline 
$-(j+\frac{1}{2})$&
$-\kappa$&
$j+\frac{1}{2}=-\kappa$ \tabularnewline
\hline 
$(j+\frac{1}{2})$&
$\kappa+1$&
$j-\frac{1}{2}=\kappa-1$ \tabularnewline
\hline
\end{tabular}\label{tab-quantum}
\end{table}

\subsubsection*{Case (i) $\kappa=-(j+1/2)$ :}

From table \ref{tab-quantum}, it is easy to see that for this case,
$n_{\kappa}=l_{B}=-\kappa$. The integral in Eq.(\ref{**}) becomes

\begin{eqnarray}
g_{\kappa i}^{S}(p) & = & \frac{N_{\kappa i}^{S}N_{\kappa i}^{L}}{\sqrt{p}}\int_{0}^{\infty}(-2\alpha_{i})r^{(n_{\kappa}+3/2)}e^{-\alpha_{i}r^{2}}J_{n_{\kappa}+1/2}(pr)\, dr,\nonumber \\
 & = & -N_{\kappa i}^{S}N_{\kappa i}^{L}\frac{p^{n_{\kappa}}}{(2\alpha_{i})^{(n_{\kappa}+1/2)}}e^{-p^{2}/4\alpha_{i}}.\label{case1}\end{eqnarray}

\subsubsection*{Case (ii) $\kappa=(j+1/2)$ : }

For this case, $l_{B}=n_{\kappa}-2=\kappa-1$, which upon substitution
in Eq.(\ref{**}) yields

\begin{eqnarray}
g_{\kappa i}^{S}(p) & = & \frac{N_{\kappa i}^{S}N_{\kappa i}^{L}}{\sqrt{p}}\int_{0}^{\infty}\left[(2n_{\kappa}-1)r^{(n_{\kappa}-1/2)}-2\alpha_{i}r^{(n_{\kappa}+3/2)}\right]e^{-\alpha_{i}r^{2}}J_{n_{\kappa}-3/2}(pr)\, dr.\label{eq-gsp}\end{eqnarray}

Next we use the result\cite{abram+steg} 

\begin{eqnarray}
\int_{0}^{\infty}x^{\mu}e^{-\alpha x^{2}}J_{\nu}(\beta x)\, dx & = & \frac{\beta^{\nu}\Gamma\left(\frac{\nu}{2}+\frac{\mu}{2}+\frac{1}{2}\right)}{2^{\nu+1}\alpha^{\frac{1}{2}(\mu+\nu+1)}\Gamma(\nu+1)}\Phi\left(\frac{\nu+\mu+1}{2},\nu+1,-\frac{\beta^{2}}{4\alpha}\right)\nonumber \\
 &  & \mbox{for }Re(\alpha)>0,Re(\mu+\nu)>-1,\label{eq-ppp}\end{eqnarray}
 where $\Phi(a,b,z)$ is the confluent hypergeometric function, in
Eq. (\ref{eq-gsp}), and after some simplifications obtain

\begin{eqnarray}
g_{\kappa i}^{S}(p) & = & \frac{N_{\kappa i}^{S}N_{\kappa i}^{L}}{\sqrt{p}}\frac{p^{(n_{\kappa}-3/2)}}{2^{(n_{\kappa}-3/2)}{\alpha_{i}}^{(n_{\kappa}-1/2)}}\left[\left(n_{\kappa}-\frac{1}{2}\right)\Phi\left(n_{\kappa}-\frac{1}{2},n_{\kappa}-\frac{1}{2},-\frac{p^{2}}{4\alpha_{i}}\right)\right.\nonumber \\
 &  & \left.-\left(n_{\kappa}-\frac{1}{2}\right)\Phi\left(n_{\kappa}+\frac{1}{2},n_{\kappa}-\frac{1}{2},-\frac{p^{2}}{4\alpha_{i}}\right)\right]\label{eq-big}\end{eqnarray}
 Next, we use the following two identities involving the confluent
hypergeometric functions\cite{abram+steg}

\begin{equation}
a\Phi(a+1,b,z)=(z+2a-b)\Phi(a,b,z)+(b-a)\Phi(a-1,b,z),\label{hypergeo1}\end{equation}
and \begin{equation}
\Phi(a,a,z)=e^{z},\label{hypergeo2}\end{equation}
to obtain the following simple expression from Eq. (\ref{eq-big})\begin{equation}
g_{\kappa i}^{S}(p)=N_{\kappa i}^{S}N_{\kappa i}^{L}\frac{p^{n_{\kappa}}}{(2\alpha_{i})^{(n_{\kappa}+1/2)}}e^{-p^{2}/4\alpha_{i}}.\label{case2}\end{equation}

Comparing the results of two cases (\ref{case1}) and (\ref{case2}),
we find that they only differ by a sign, and hence when substituted
in the expression for $J_{ij}^{S;\kappa}(q)$ in Eq.(\ref{js}) yield
the same result

\[
J_{ij}^{S;\kappa}(q)=\frac{1}{2}\int_{q}^{\infty}(N_{\kappa i}^{S})(N_{\kappa j}^{S})(N_{\kappa i}^{L})(N_{\kappa j}^{L})\frac{p^{2n_{\kappa}+1}}{(4\alpha_{i}\alpha_{j})^{n_{\kappa}+1/2}}e^{-p^{2}/4\alpha_{ij}}\, dp,\]
 where $\alpha_{ij}=\frac{\alpha_{i}\alpha_{j}}{\alpha_{i}+\alpha_{j}}$
. The above integral can be evaluated in exactly the same way as was
done before for the large component (\emph{cf.} \ref{jl-final}),
to yield the final expression for the Compton profile matrix element\begin{equation}
J_{ij}^{S;\kappa}(q)=\frac{(N_{\kappa i}^{S})(N_{\kappa j}^{S})(N_{\kappa}^{L}i)(N_{\kappa j}^{L})}{4}\frac{(4\alpha_{ij})^{l_{A}+2}}{(4\alpha_{i}\alpha_{j})^{l_{A}+3/2}}\Gamma(l_{A}+2,q_{t}),\label{js_final}\end{equation}
 where $q_{t}=\frac{q^{2}}{4\alpha_{ij}}$, and the incomplete gamma
function is defined in Eq. (\ref{igamma}). Finally, the large and
small components of the CP of an orbital can be computed in terms
of these matrix elements, as

\begin{equation}
J_{n\kappa}^{L}(q)=\sum_{i,j}C_{\kappa i}^{L}C_{\kappa j}^{L}J_{ij}^{L;\kappa},\end{equation}
 \begin{equation}
J_{n\kappa}^{S}(q)=\sum_{i,j}C_{\kappa i}^{S}C_{\kappa j}^{S}J_{ij}^{S;\kappa}.\end{equation}

It is these formulas derived here which have been numerically implemented
in our computer program COMPTON\cite{cp-program} aimed at calculating
relativistic atomic CPs.

\end{document}